\documentclass[preprint,nofootinbib,superscriptaddress,amssymb]{revtex4}
\usepackage{amsmath} 
\usepackage{feynmp}	
\usepackage{verbatim} 
\usepackage{slashed} 
\usepackage{graphicx} 
\usepackage{subfigure} 

\begin{document}	
\unitlength = 1mm 
\renewcommand{\thesubfigure}{(\arabic{subfigure})}

\begin{fmffile}{graphs}

\title{Quantum Equivalence Principle Violations \\ in Scalar-Tensor Theories}

\author{Cristian Armendariz-Picon}
\affiliation{Department of Physics, Syracuse University, Syracuse, NY 13244-1130, USA}

\author{Riccardo Penco}
 \affiliation{Department of Physics, Syracuse University, Syracuse, NY 13244-1130, USA}
 
\begin{abstract}
We study the  equivalence principle and its violations by quantum effects in scalar-tensor theories that admit a conformal frame in which matter only couples to the spacetime metric.  These theories possess  Ward identities that guarantee the validity of the weak equivalence principle to all orders in the matter coupling constants. These Ward identities originate from a broken Weyl symmetry  under which the scalar field transforms by a shift, and from the symmetry required to couple a massless spin two particle to matter (diffeomorphism invariance). But the same identities also predict violations of the weak equivalence principle relatively suppressed by at least two powers of the gravitational couplings, and imply that quantum corrections do not preserve the structure of the action of these theories.  We illustrate our analysis with a set of specific examples for spin zero and spin half matter fields that show why matter couplings do respect the equivalence principle, and how the couplings to the gravitational scalar lead to the weak equivalence principle violations predicted by the Ward identities. 

\end{abstract}

\maketitle

\section{Introduction}

Einstein based the development of general relativity on two  pillars: general covariance and the equivalence principle. Since then, physicists have often wondered whether there are any alternatives to general relativity, which, while preserving its theoretical framework and phenomenological successes  also avoid some of the shortcomings sometimes attributed  to it. Among the phenomenological successes of general relativity, the  equivalence principle---the proportionality of inertial and gravitational mass---is the most accurately tested and constrained one. Indeed, experiments at the University of Washington limit the relative difference in acceleration towards the earth of two  test spheres of different atomic compositions to be less than one part in $10^{12}$ \cite{Schlamminger:2007ht}. Therefore, any putative alternative theory of gravitation has to pass the significant hurdle of the  equivalence principle. 

Arguably, the simplest way to modify general relativity is to add a scalar field to the gravitational sector. Since gravitation is a long-ranged interaction, such a scalar  would have to be sufficiently light to be considered  part of the gravitational field. Whereas it is straightforward to include such a scalar field  while preserving diffeomorphism invariance, the most general diffeomorphism-invariant theory with a light scalar would generically lead to strong violations of the equivalence principle \cite{Damour:2010rm}. There is nevertheless a subclass of scalar-tensor theories which respect the weak form of the equivalence principle, at least at tree level, and thus provides a natural class of phenomenologically viable alternatives to general relativity. (As shown by Nordtvedt \cite{Nordtvedt:1968qs}, theories in this class do  violate the strong equivalence principle, although these violations are negligible in laboratory-sized experiments.) The first such theory  was proposed by Pascual Jordan \cite{Jordan:1959eg}, after criticism by Fierz \cite{Fierz:1956zz} of an earlier proposal of the former \cite{Jordan:1955}. Essentially the same theory  was later revived  by Brans and Dicke \cite{Brans:1961sx}, whose names are usually associated with the class of scalar-tensor theories we study here. Further extensions and generalizations within this class  were later considered by different authors \cite{Wagoner:1970vr}. 

What distinguishes these weak equivalence principle-preserving scalar-tensor theories is the existence of a formulation of the theory---a conformal frame---in which the scalar field only couples to gravity (at tree level). It follows then, by construction, that these theories preserve the weak equivalence principle classically,  since their matter sector is the same as that of general relativity. Of course, the question is what happens to the  equivalence principle when quantum fluctuations are turned on, and, more generally, whether quantum corrections preserve the structure of this subclass of scalar-tensor theories. This is not just a purely academic question, because even Planck-suppressed interactions eventually generated by loops would lead to departures from the weak equivalence principle that are experimentally ruled out. The question is most conveniently addressed in the Einstein conformal frame of these theories, in which the propagators of the graviton and the scalar are diagonal.  Although in this frame the scalar couples directly to matter, it is easy to check that the equivalence principle is preserved at tree level. However, because the field couples directly to matter, it is hard to see why quantum corrections would not lead to violations of the equivalence principle. 

The impact  of quantum corrections on the equivalence principle has been the subject of a small but interesting debate in the literature. In the first article on the topic we were able to find, Fujii  argued that quantum corrections should violate the equivalence principle \cite{Fujii:1994cn}. He explicitly calculated  one-loop quantum corrections to the vertex for scalar emission by a photon, with matter fields running inside the loop, and  argued that the latter do  seem to violate the equivalence principle. But somewhat later the same author realized that this purported violation disappears if one employs dimensional regularization instead of a  cut-off \cite{Fujii:1996td}. Unaware of these results, Cho also argued that the equivalence principle should be violated in scalar-tensor theories \cite{Cho:1994qv}, though some of his arguments seemed to be in conflict with the  explicit calculation performed by Fujii in \cite{Fujii:1996td}. Up to that point, whether or why quantum corrections preserve  the equivalence principle remained unclear, to say the least.  Recently, Hui and Nicolis  have shed more light on the issue by providing explicit  examples for \emph{massive} fields showing that   matter loops do not lead to violations of the weak equivalence principle in scalar-tensor theories  \cite{Hui:2010dn}. They  argue that this is due  to the linear coupling of the scalar to the trace of the energy momentum tensor: Because the energy-momentum tensor is conserved, the scalar couples to a charge density given by the time-time component of the energy-momentum tensor, which they identify with the mass density. 

In this article we extend these arguments further. As we shall see, the equivalence principle in scalar-tensor theories has a two-fold origin: A broken  Weyl symmetry that relates the couplings of the scalar to those of the graviton, and diffeomorphism invariance, which significantly constrains the couplings of the graviton (and demands in particular that the latter couple to a conserved quantity, the energy-momentum tensor.) Diffeomorphism invariance  implies that in the limit of zero  momentum transfer the vertex for graviton emission by matter---the gravitational mass---has to be proportional to the inertial mass \cite{Weinberg:1964ew,BroutEnglert:1966,DeWitt:1967uc}, and it is the broken Weyl symmetry what makes the couplings to the scalar inherit that property. Moreover, because this Weyl symmetry is broken, there is a corresponding Ward identity for the broken symmetry that exactly predicts the size of those quantum corrections that violate the equivalence principle:  They have to be proportional to three inverse powers of the gravitational couplings.

\section{Formalism}

\subsection{Action Principle}
The scalar-tensor theories we are about to study are characterized by the existence of a  conformal frame, the Jordan frame, in which bosonic  matter is minimally coupled to the spacetime metric.  Out of all possible scalar-tensor theories, this restriction singles out a very specific class of theories in which the weak equivalence principle holds, at least classically.

By definition, the gravitational sector of any scalar-tensor theory consists of a scalar $\phi$ and a rank two symmetric tensor $g_{\mu\nu}$, the metric. Our universe contains fermionic fields however, and this conventional formulation has to be replaced by one in terms of the scalar $\phi$ and the vierbein $e_\mu{}^a$ (see \cite{Carroll:2004st} for a review.) In this language, the scalar-tensor theories we consider here have an action functional
\begin{equation}\label{eq:S JF}
	S_J=\int d^d x \det e\left[F(\phi)R -G(\phi)\partial_\mu \phi \partial^\mu \phi-W(\phi)\right] +S^J_M[e_\mu{}^a,\psi_\alpha],
\end{equation}
where the index $J$ denotes Jordan frame quantities, $\psi_\alpha$ is a set of matter fields and $R$ is the Ricci scalar associated with the metric
\begin{equation}\label{eq:g}
	g_{\mu\nu}=\eta_{ab} e_\mu{}^a e_\nu{}^b.
\end{equation}
Note that we work in an arbitrary number of dimensions $d$, and that matter is now minimally coupled to the vierbein field, which is what singles out the class of theories we consider in this article. To some extent the dynamics of the gravitational sector are unimportant; our considerations can be easily generalized to  even more general forms of the gravitational sector of the action. 

The action (\ref{eq:S JF}) is invariant under two symmetry groups: diffeomorphisms and local Lorentz transformations. Because any spacetime tensor can be converted into a diffeomorphism scalar by contraction with the vierbein, we can assume that all matter fields are diffeomorphism scalars. In that case, under infinitesimal diffeomorphisms $x^\mu\to x'^\mu=x^\mu+\xi^\mu(x)$ the fields of the theory transform according to
\begin{subequations}\label{eq:diffs} 
\begin{align}
	&e_\mu{}^a \to e'_\mu{}^a=
	e_\mu{}^a+\Delta e_\mu{}^a,\quad 
	& &\Delta e_\mu{}^a=-\xi^\nu \partial_\nu e_\mu{}^a- e_\nu{}^a \partial_\mu \xi^\nu, \\
	&\phi\to\phi'=\phi+\Delta\phi,\quad
	& &\Delta\phi=-\xi^\mu \partial_\mu \phi, \\
	&\psi_\alpha\to\psi_\alpha'=\psi_\alpha+\Delta\psi_\alpha,\quad
	& &\Delta\psi_\alpha=-\xi^\mu \partial_\mu \psi_\alpha.
\end{align}
\end{subequations}
Under local Lorentz transformations $\Lambda(x)\in SO(1,3)$ the different fields transform in the corresponding representation of the Lorentz group,
\begin{subequations}
\begin{align}
	&e_\mu{}^a\to e'_\mu{}^a=\Lambda^a{}_b \, e_\mu{}^b,\\
	&\phi\to\phi'=\phi,\\
	&\psi_\alpha\to\psi_\alpha'=D(\Lambda)_\alpha{}^\beta\psi_\beta,
\end{align}
\end{subequations}
where $D$ is the linear representation of the Lorentz group under which the matter fields transform. 

Our goal is to investigate the gravitational interactions experienced by the different matter fields. In the quantum theory these interactions  are mediated by the interchange of gravitons and scalar particles. However, in the action (\ref{eq:S JF}) the graviton and scalar propagators are typically not diagonal. Hence, it is convenient and customary to introduce a new set of variables in terms of which the propagators become diagonal. This set of new variables define what is usually known as the Einstein frame, in which the action reads
\begin{subequations}\label{eq:S EF}
\begin{equation}\label{eq:S EF M}
	S_E=S_{EH}[e_\mu{}^a]+S_\phi[e_\mu{}^a,\phi]
	+S^E_M[f(\phi/M)e_\mu{}^a,\psi_\alpha],
\end{equation}
where
\begin{equation}\label{eq:S EH phi}
	S_{EH}=\int d^d x \det e\, \frac{M_P^2}{2}R, \quad 
	S_\phi=\int d^d x \det e \left[-\frac{1}{2}\partial_\mu \phi \partial^\mu \phi-V(\phi)\right].
\end{equation}
\end{subequations}
Of course, the choice of conformal frame is a matter of convenience, and both (\ref{eq:S JF}) and (\ref{eq:S EF}) are physically equivalent, as recognized early on by Dicke  \cite{Dicke:1961gz} (in the quantum theory, the equivalence follows from the invariance of $S$-matrix elements under field redefinitions.) For convenience and simplicity we take however (\ref{eq:S EF}) as the starting point of our considerations. We also  assume that the equations of motion admit a solution with constant value $\bar{\phi}$ of the scalar field, which for simplicity we take to be $\bar{\phi}=0$, and  at this minimum we define
\begin{equation}
	m_\phi^2\equiv \frac{d^2 V}{d\phi^2}\Bigg|_{\bar\phi=0}.
\end{equation}
 If both $f(0)$ and $f'(0)$ differ from zero, we may assume without loss of generality the normalization conditions
\begin{equation}\label{eq:f normalization}
	f(0)=1,\quad  f'(0)=1.
\end{equation}
 Note that in $d$ spacetime dimensions, $M_P$ and $M$ do not have mass dimension  one. Instead they have  the same  mass dimension as the scalar and the graviton. 
 
\subsection{The Weak Equivalence Principle}
\label{sec:The Weak Equivalence Principle}

\begin{figure}
\subfigure[]
{
\begin{fmfgraph*}(50,30) 
\fmfleft{i1,i2} \fmfright{o1,o2} 
\fmflabel{$p_A$}{i2}\fmflabel{$p'_A$}{o2}
\fmflabel{$p_B$}{i1}\fmflabel{$p'_B$}{o1}
\fmf{fermion}{i1,v1,o1} \fmf{fermion}{i2,v2,o2} 
\fmf{dashes,label=$q$}{v1,v2}
\end{fmfgraph*}
}
\hspace{2cm}
\subfigure[]
{
\begin{fmfgraph*}(50,30) 
\fmfleft{i1,i2} \fmfright{o1,o2} 
\fmflabel{$p_A$}{i2}\fmflabel{$p'_A$}{o2}
\fmflabel{$p_B$}{i1}\fmflabel{$p'_B$}{o1}
\fmf{fermion}{i1,v1,o1} \fmf{fermion}{i2,v2,o2} 
\fmf{dbl_wiggly,label=$q$}{v1,v2}
\end{fmfgraph*}
}

\caption{Scalar and graviton exchange between two different matter  species. Continuous lines denote matter fields (bosonic or fermionic), dashed lines label the scalar $\phi$, and wiggly lines label the graviton.}
\label{fig:tree scalar exchange}
\end{figure}
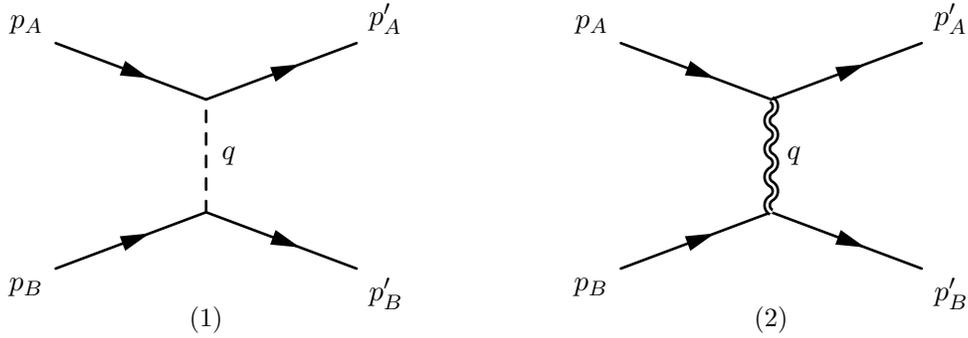

Recall that the weak equivalence principle states that in a gravitational field all neutral test bodies fall with the same acceleration, or, more simply, that gravitational and inertial mass are proportional to each other.  To see how the equivalence principle emerges in the classical theory defined by the action (\ref{eq:S EF M}), consider the tree-level diagram in figure \ref{fig:tree scalar exchange}.1, in which two different matter particles scatter through scalar exchange on a Minkowski background.  The amplitude of the diagram in  figure \ref{fig:tree scalar exchange}.1 is\footnote{We mostly follow the conventions of \cite{Weinberg:1995mt}. In these conventions, the propagator carries a factor of $(2\pi)^{-d}$, each external line contributes a factor of $(2 \pi)^{-(d-1)/2}$, and the relation between  S-matrix elements and the amplitudes $\mathcal{M}$ for an initial state $i$ and a final state $f$ is $S_{fi} = \delta_{fi} - 2 \pi i \delta (p_f - p_i) \mathcal{M}_{fi}$. See the next subsections for additional information on our conventions for vertices and propagators.}
\begin{equation}\label{eq:tree M}
	\mathcal{M}_\phi=  - \frac{1}{(2\pi)^{3d-1}} [u^\dagger_\beta (p'_A)  \gamma_\phi^{\beta\alpha} u_\alpha (p_A)] 
	\,\frac{1}{q^2+m_\phi^2}\,
	[u^\dagger_\beta (p'_B) \gamma_\phi^{\beta\alpha} u_\alpha (p_B)],
\end{equation}
where $\gamma^{\beta\alpha}_\phi$ is the tree-level amplitude for scalar emission by matter, the $u_\alpha(p)$ are the appropriate mode functions for the external particles, $q\equiv p_A'-p_A$ is the momentum transfer, and $(q^2+m_\phi^2)^{-1}$ is the scalar propagator. We are interested here in the potential energy between two static bodies, that is, on a scalar whose four-momentum $q^\mu$ approaches zero: $p_A'\to p_A,\,  p_B'\to p_B$.

Inspection of the way $\phi$ enters the action (\ref{eq:S EF M}) reveals that in flat spacetime the scalar vertex $\gamma_\phi$ is related to the vertex for graviton  emission $(\gamma_h)^{\mu\nu}$ by
\begin{equation}\label{eq:phi h}
	M \gamma_\phi^{\beta\alpha}= 2M_P   (\gamma_h^{\beta\alpha})^\mu{}_\mu,
\end{equation}
where we have used equation (\ref{eq:f normalization}). The graviton vertex $\gamma_h$ is proportional to the quadratic component of the energy momentum tensor in flat space, so equation (\ref{eq:phi h}) is just roughly the statement that the scalar couples to the trace of the energy-momentum tensor (the factor of two stems from the identification of the vierbein as ``half" a graviton).  As we shall see, it follows from diffeomorphism invariance alone that in momentum space, and in the limit of zero momentum transfer, this tree-level graviton vertex has to be of the form
\begin{equation}\label{eq:Ward tree q=0}
	2M_P (\gamma^{\beta\alpha}_h)^\mu{}^\nu=
	\pi^{\beta\alpha}(p)\, \eta^\mu{}^\nu-
	p^{(\mu} \frac{\partial \pi^{\beta\alpha}}{\partial p_{\nu)}},
\end{equation}
where a parenthesis next to an index denotes symmetrization, $p^\mu$ is the momentum of matter, and $\pi^{\alpha\beta}$ is the tree-level self-energy, that is, minus the inverse of the tree-level propagator.  The reader can easily verify this relation in the cases of a scalar, a spin half fermion and spin one vector. Hence, because of equation (\ref{eq:phi h}), an analogous relation applies for the amplitude for scalar emission,
\begin{equation}\label{eq:Ward phi tree}
	M \gamma_\phi^{\beta\alpha}=d\,\pi^{\beta\alpha}-p^\mu \frac{\partial \pi^{\beta\alpha}}{\partial p^\mu},
\end{equation}
which, again, can be checked independently for scalars, spinors and vectors. On shell, the self-energy $\pi$ vanishes by definition. Contracting then equation (\ref{eq:Ward phi tree}) with the appropriate mode functions  we  find for all three types of matter fields that, on shell,
\begin{equation}\label{eq:Ward tree c}
	u^\dagger_\beta \gamma^{\beta\alpha}_\phi u_\alpha =\frac{(2\pi)^d}{M} \frac{p^2}{p^0}
	=-\frac{(2\pi)^d}{M} \frac{m_I^2}{p^0},
\end{equation}
where $m_I$ is the inertial mass of the particle, defined to be the value of $-p^2$ at the zero of the self-energy, and we have also used that for free fields of arbitrary spin
\begin{equation}\label{eq:identity Lorentz}
	u_\beta^\dag  \frac{\partial\pi^{\beta\alpha}}{\partial p_\mu}u_\alpha
	= - (2 \pi)^d \, \frac{p^\mu}{p^0}.
\end{equation}
In particular, note that equation (\ref{eq:Ward tree c}) implies that massless particles do not couple to the scalar at tree level, even if the field Lagrangian is not conformally invariant, as happens for instance for a massless scalar. Hence, the scalar interaction does not contribute to the bending of light, and the experimental constraints on the Eddington parameter $\gamma$  thus demand that the scalar interaction be much weaker than gravity, $M_P\ll M$ \cite{Will:2005va}. Finally, substituting equation (\ref{eq:Ward tree c}) into (\ref{eq:tree M}) and taking the limit of non-relativistic massive particles, $p^0\approx m_I$, we arrive at
\begin{equation}\label{eq:Mphi}
	\mathcal{M}_\phi= - \frac{1}{(2 \pi)^{d-1}} \, \frac{m_A m_B}{M^2}\frac{1}{q^2+m_\phi^2},
\end{equation}
where $m_A$ and $m_B$ are, respectively, the inertial masses of particles $A$ and $B$.  

As we mentioned above, we want to calculate the potential energy  for two static bodies, at fixed spatial distance $\vec{r}$ in $d=4$ spacetime dimensions.  To this end, we simply need to Fourier transform the non-relativistic limit of the scattering amplitude (\ref{eq:Mphi}) back to real space. Since in the non-relativistic limit $q^2=\vec{q}\,^2$, we obtain
\begin{equation}\label{eq:V}
	V(\vec{r})\equiv \int d^3q \, e^{i \vec{q}\cdot \vec{r}}\, \mathcal{M}_\phi(\vec{q}) = - \frac{m_A m_B}{M^2}\frac{e^{-m_\phi r}}{4 \pi r}.
\end{equation}
Hence, the force mediated by the scalar $\phi$ is proportional to the inertial mass, with a proportionality factor $1/M^2$ that is  universal:  The scalar interaction respects the weak equivalence principle.

Of course, if we calculate the potential energy due to graviton exchange, we also find that the latter respects the weak equivalence principle. As before, on shell, using equations (\ref{eq:Ward tree q=0}) and (\ref{eq:identity Lorentz}) we find
\begin{equation}
	u^\dagger_\beta (\gamma^{\beta\alpha}_h)^{\mu\nu} u_\alpha=
	\frac{(2\pi)^d}{2M_P} \, \frac{p^{\mu} p^{\nu}}{p^0}.
\end{equation}
Since with our conventions the graviton  propagator  in $d$ spacetime dimensions (say, in de~Donder gauge)  is
\begin{equation}
	i [\pi_h^{-1}(q)]_{\mu\nu,\rho\sigma} = \frac{- 4i }{(2\pi)^d q^2} \left\{  \frac{1}{2} \left( \eta_{\mu\rho}\eta_{\nu\sigma} +\eta_{\mu\sigma}\eta_{\nu\rho}\right)  - \frac{1}{d-2} \, \eta_{\mu\nu}\eta_{\rho\sigma}  \right\}
\end{equation}
we obtain in the non-relativistic limit that the amplitude associated with the diagram in figure \ref{fig:tree scalar exchange}.2 in $d=4$ is
\begin{equation}
	\mathcal{M}_h= - \frac{m_A m_B}{2 M_P^2}\frac{1}{q^2}.
\end{equation}
Again, the amplitude is proportional to the inertial masses of both particles, with a proportionality constant $1/M_P^2$ that is universal. The origin of this result is the tree-level Ward-Takahashi identity (\ref{eq:Ward tree q=0}). The latter relates emission of a graviton---the gravitational mass---to the self-energy of matter---the inertial mass. It just so happens that, due to the structure of the matter action in (\ref{eq:S EF M}), the scalar couplings ``inherit" this Ward identity, ultimately leading to the preservation of the weak equivalence principle in the scalar sector (at tree level). We explore whether these features survive in the quantum theory next. 

\subsection{Quantization}
 
For the purpose of quantization, it shall prove to be useful to work with the quantum effective action $\Gamma$, the sum of all one-particle-irreducible (1PI) diagrams with a given number of external lines. In order to calculate the effective action, we expand the fields in quantum fluctuations around  a given (but arbitrary)  background.   We thus write 
\begin{subequations}
 \begin{eqnarray}
 	e_\mu{}^a&=&\bar{e}_\mu{}^a+M_P^{-1}  \delta e_\mu{}^a, \label{eq:delta e}\\
	\phi&=&\bar{\phi}+\delta\phi, \\
	\psi_\alpha&=&\bar{\psi}_\alpha+\delta\psi_\alpha,
 \end{eqnarray}
\end{subequations}
where overbars denote background values, and deltas quantum fluctuations. Plugging equation (\ref{eq:delta e}) into (\ref{eq:g}) we find
 \begin{equation}\label{eq:metric fluctuations}
	 g_{\mu\nu}\equiv \bar{g}_{\mu\nu}+M_P^{-1} \, h_{\mu\nu}+\mathcal{O}(M_P^{-2}), \quad \text{with} \quad 
	 \bar{g}_{\mu\nu}=\eta_{ab}\bar{e}_\mu{}^a\bar{e}_\nu{}^b, \quad
	 h_{\mu\nu}=\delta e_{\mu\nu}+\delta e_{\nu\mu},
 \end{equation}
and $\delta e_{\mu\nu}\equiv  \bar{e}_\nu{}_a \delta e_\mu{}^a$ (note that the location of the vierbein indices is important.) Hence, the symmetric part of the vierbein fluctuations, $h_{\mu\nu}$, is the graviton field; its antisymmetric part $a_{\mu\nu}\equiv \delta e_{\mu\nu}-\delta e_{\nu\mu}$ is non-dynamical \cite{Deser:1974cy}. It follows then by definition that\footnote{Roughly speaking, just as we think of the vierbein as the square root of the metric, we can think of a vierbein fluctuation as half a graviton.}
 \begin{equation}\label{eq:h a}
 	\delta e_{\mu\nu}=\frac{h_{\mu\nu}}{2}+\frac{a_{\mu\nu}}{2}.
 \end{equation}
 
As in any non-abelian gauge theory, we quantize the theory defined by (\ref{eq:S EF})  using the functional integral formalism.  Because the action (\ref{eq:S EF}) is invariant under two groups of local symmetries (diffeomorphisms and Lorentz transformations), we need to fix both gauges and introduce the corresponding ghost fields. Hence, our total action becomes 
\begin{equation}\label{eq:S tot}
	S_\mathrm{tot}=S_E+S_{GF}+S_G,
\end{equation}
where $S_E$ is given in equation (\ref{eq:S EH phi}), $S_{GF}$ is the gauge-fixing term, and $S_G$ the action for the ghosts. In the  background field method, the gauge fixing term is such that the total action $S_\mathrm{tot}$ in equation (\ref{eq:S tot}) is invariant under a set of symmetries in which the background fields transform like the fields themselves, that is, under equations (\ref{eq:diffs}). For concreteness, and following \cite{Deser:1974cy}, we impose  the de~Donder (harmonic) gauge condition to fix the diffeomorphism gauge, and an algebraic term to fix the Lorentz frame, 
\begin{equation}
	S_{GF}=-\frac{1}{4}\int d^d x \det \bar{e}\, \left[ \bar{g}^{\mu\nu}\, 
	\left(\bar\nabla_\rho h^\rho{}_\mu-\frac{1}{2}\bar{\nabla}_\mu h^\rho{}_\rho\right)
	\left(\bar\nabla_\rho h^\rho{}_\nu-\frac{1}{2}\bar{\nabla}_\nu h^\rho{}_\rho\right)
	+\bar{g}^{\mu\rho}\bar{g}^{\nu\sigma} \frac{a_{\mu\nu}a_{\rho\sigma}}{2M_P^2}\right].
\end{equation}
With this choice of gauge fixing, the action for the diffeomorphism  ghosts $\zeta^\mu$ and the Lorentz ghosts $\theta^{\mu\nu}$ becomes
\begin{equation}
	S_{G}=-\frac{1}{\sqrt{2}}\int \det \bar{e} 
	\left[\zeta^{\dag\mu} 
	\left(\bar{g}_{\mu\nu}\bar\Box-\bar{R}_{\mu\nu}\right)
	\zeta^\nu+\frac{M_P^2}{2}\bar{g}_{\mu\rho}\bar{g}_{\nu\sigma}
	\theta^\dag{}^{\mu\nu}\theta^{\rho\sigma}\right] .
\end{equation}
We employ dimensional regularization, which preserves the gauge symmetries of the theory while rendering the theory finite. The effective action is the path integral over these fluctuations,
with the prescribed values of the background fields kept fixed, 
\begin{equation}\label{eq:Gamma}
	\exp(i\Gamma[\bar{e}_\mu{}^a, \bar{\phi}, \bar{\psi}_\alpha])\equiv
	\int_{1PI}
	\mathcal{D}\delta e \, \mathcal{D}\delta \phi\, \mathcal{D}\delta\psi \,
	\mathcal{D}\zeta\, \mathcal{D}\theta \,
	\exp[iS_\mathrm{tot}].
\end{equation}
The integral is restricted to run only  over all one-particle-irreducible vacuum diagrams. The end result of this construction is that the effective action remains invariant under diffeomorphisms and Lorentz transformations, even though these symmetries had to be broken to define the path integral. 

\subsection{Gravitational Interactions}
\label{sec:Gravitational Interactions}

\begin{figure}
\subfigure[]
{
\begin{fmfgraph*}(40,30) 
\fmfleft{i1,i2} \fmfright{o1,o2}
\fmflabel{$p_A$}{i2}\fmflabel{$p'_A$}{o2}
\fmflabel{$p_B$}{i1}\fmflabel{$p'_B$}{o1} 
\fmfblob{.15w}{v1}
\fmfblob{.15w}{v2} 
\fmfblob{.1w}{v3}
\fmf{fermion}{i1,v1,o1} \fmf{fermion}{i2,v2,o2} 
\fmf{dashes}{v1,v3,v2}
\end{fmfgraph*}
}
\hspace{2cm}
\subfigure[]
{
\begin{fmfgraph*}(40,30) 
\fmfleft{i1,i2} \fmfright{o1,o2} 
\fmflabel{$p_A$}{i2}\fmflabel{$p'_A$}{o2}
\fmflabel{$p_B$}{i1}\fmflabel{$p'_B$}{o1}
\fmfblob{.15w}{v1}
\fmfblob{.15w}{v2} 
\fmfblob{.1w}{v3}
\fmf{fermion}{i1,v1,o1} \fmf{fermion}{i2,v2,o2} 
\fmf{dbl_wiggly}{v1,v3,v2}
\end{fmfgraph*}
}
\caption{A light scalar (dashed) and a massless graviton mediate long-ranged interactions through the interchange of a single quantum. Each blob in a vertex represents the sum of all one-particle-irreducible diagrams (1PI) with the corresponding number of external lines, and external propagators stripped off. Each blob with two external lines represents the full propagator, the sum of all (connected) diagrams with the corresponding type of external lines.}
\label{fig:long range}
\end{figure}
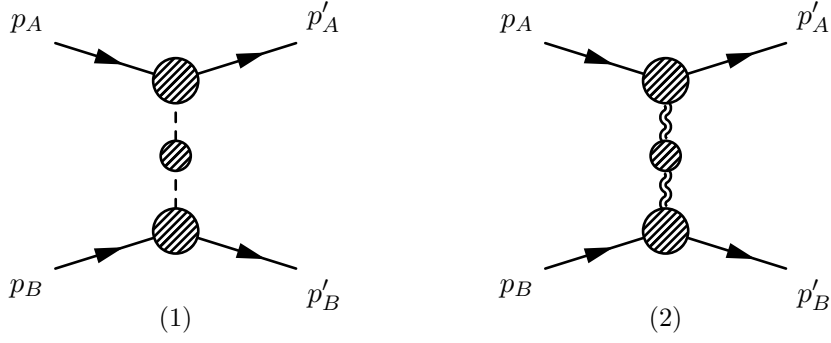

Consider now the scattering of two distinguishable particles described by the matter fields $\psi_\alpha$ (and their adjoints $\psi_\alpha^\dag$ when appropriate). Restricting ourselves to interactions mediated by the vierbein and the scalar, these are determined by the two diagrams in figure \ref{fig:long range}, the counterparts of the two tree-level diagrams of figure \ref{fig:tree scalar exchange}. In real space, the 1PI vertices are given by functional derivatives of the quantum effective action evaluated at vanishing field fluctuations. In particular, in view of (\ref{eq:metric fluctuations}) and (\ref{eq:h a}), the irreducible vertices for emission of a graviton and a scalar by matter are
\begin{equation}\label{eq:vertices}
(\Gamma_h^{\beta\alpha})^{\mu\nu}{}(z; y, x)\equiv 
	\frac{1}{2M_P}	\frac{\delta^3\Gamma}{\delta\bar{\psi}_\alpha(x) \delta\bar{\psi}^\dag_\beta(y)\delta\bar{e}_{(\mu}{}^a(z)}\bar{e}^{\nu) a}(z)
 , \quad
\Gamma_\phi^{\alpha\beta}(z;y, x)\equiv 
	\frac{\delta^3\Gamma}{ \delta\bar{\psi}_\alpha(x)\delta\bar{\psi}^\dag_\beta(y)\delta\bar{\phi}(z)},
\end{equation} 
while the self-energies of the graviton and the scalar (minus the inverse of their propagator) are given by\footnote{Because the effective action is diffeomorphism invariant by construction, we need to add to it an additional gauge fixing term to define the graviton propagator \cite{Abbott:1983zw}.} 
\begin{equation}\label{eq:self-energies}
	(\Pi_h)^{\mu\nu, \rho\sigma}(y, x)\equiv 
	\bar{e}^{(\rho a}(x)
	\frac{\delta^2\Gamma}{\delta\bar{e}_{\sigma)}{}^a(x)\delta\bar{e}_{(\mu}{}^b(y)}
	 \bar{e}^{\nu b}(y),
	 \quad 
	\Pi_\phi(y,x) \equiv 
	\frac{\delta^2\Gamma}{ \delta\bar{\phi}(x)\delta\bar{\phi}(y)}.
\end{equation}
These functional derivatives are evaluated in a Minkowski spacetime background with vanishing scalar and matter fields,
\begin{equation}\label{eq:background}
	\bar{\phi}=0,\quad \bar{\psi}_\alpha=0,\quad 
	\bar{e}_\mu{}^a=\delta_\mu{}^a,
\end{equation}
though we do not make this  explicit (it should be clear from the context.) If, aside from the vierbein, the background does not contain any Lorentz vectors, the variational derivative $\delta^2 \Gamma/(\delta \bar{e}_\mu{}^a \delta\bar{\phi})$ vanishes as a consequence of Lorentz invariance. Therefore, there is no need to consider diagrams with one incoming scalar and one outgoing graviton.  Note that the cubic vertices above describe the couplings of unrenormalized fields. To calculate physical scattering amplitudes we have to multiply these amplitudes with the appropriate wave function renormalization constants. 

Scattering amplitudes are typically calculated in momentum space, so it is convenient to work with the momentum-space vertices and self-energies defined above. In our conventions, one of the vertex momenta is incoming ($p_1$), the other two ($p_2$ and $p_3$) are outgoing, and a momentum-conserving delta function has been split off,
\begin{equation}\label{eq:momentum conventions}
	\Gamma(p_2, p_1)\,\delta(p_1 - p_2 - p_3)\equiv
		\int d^d x\, d^d y \, d^dz\,   
			\Gamma(z; y,x)  \,e^{- i p_3 z} e^{-i p_2 y} e^{i p_1 x}.
\end{equation}
In this way, the scattering amplitude is given by
\begin{equation}\label{eq:M}
\begin{split}
	\mathcal{M}&=  \frac{1}{(2\pi)^{2d-1}} \bigg[ 
	\Gamma_\phi(p'_A,p_A) \Pi^{-1}_\phi (q)\Gamma_\phi(p'_B,p_B)
	+\Gamma_h(p'_A,p_A) \Pi^{-1}_h(q) \Gamma_h(p'_B,p_B) \bigg],
\end{split}
\end{equation}
where $q \equiv p'_A-p_A=p'_B-p_B$ is the momentum transfer and the $\Gamma's$ have been contracted with the appropriate mode functions for the matter fields,
$
	\Gamma_f(p',p)\equiv 
	u^\dagger_\beta(p') \Gamma^{\beta\alpha}_f(p',p)u_\alpha(p).
$
The potential energy is determined by the values of the irreducible vertices and propagators at zero momentum transfer, $q=0$. Using the definition of potential energy and the Fourier transform in  (\ref{eq:V}),   the gravitational potential in $d=4$ becomes
\begin{equation}
	V(r)\approx-\frac{1}{2 (2\pi)^{9}}\left[\Gamma_\phi(p_A,p_A)\Gamma_\phi(p_B,p_B)\frac{Z_\phi e^{-m_\phi r}}{r}+\Gamma_h(p_A,p_A)\Gamma_h(p_B,p_B)\frac{Z_h}{2 r}
	\right],
\end{equation}
where we have used the spectral representation for the scalar propagator, and $Z_\phi$ and $Z_h$ respectively are the residues of the scalar and graviton propagators.  We assume that $m^{-1}_\phi$ is much larger than the scales $r$ under consideration, so that we can think of the force mediated by $\phi$ effectively as a long-ranged interaction (we do not consider the Chameleon mechanism here \cite{Khoury:2003aq}.) What matters for our purposes is that the potential energy is determined by the vertices for scalar and graviton emission, and, therefore, the latter dictate the fate of the equivalence principle in the quantum theory.

\section{Ward Identities}

Because the quantum effective action is invariant under diffeomorphisms, it satisfies a set of Ward-Takahashi identities that relate the full vertex for graviton emission $\Gamma_h$ to the full matter self-energy  $\Pi$, as we shall derive next.  These Ward identities are ultimately responsible for the validity of the equivalence principle in the quantum theory, as far as the couplings of matter to the graviton are concerned. 

The origin of the Ward identity for graviton emission is that the vierbein transforms non-trivially under diffeomorphisms, even for a trivial vierbein background (flat spacetime.) This is why diffeomorphism invariance strongly restricts the couplings of  matter to the graviton. In particular, it is possible to derive the weak equivalence principle in  $S$-matrix theory solely from the requirement that $S$-matrix elements be invariant under diffeomorphisms acting on the polarization vectors of the graviton  \cite{Weinberg:1964ew}.

The case of scalar emission however is quite different. The existence of a scalar field  $\phi$ coupled to matter does not require nor entail any particular symmetry. In particular, because the change in the  scalar field $\phi$ under diffeomorphisms vanishes at zero background field, diffeomorphisms have nothing to say about the couplings of the scalar to matter. This is why there is no a priori reason to expect that the couplings of the scalar field to matter respect the equivalence principle in the quantum theory. In fact they do not, as we also show further below. Nevertheless, because the scalar field only couples to matter in the combination $f(\phi/M)\,  e_\mu{}^a$, its couplings inherit the Ward identity satisfied by the graviton to all orders in the matter coupling constants.

\subsection{Graviton Emission}
Our first goal is to derive the Ward identity for graviton emission. Such an identity was proven for arbitrary bosonic matter fields by DeWitt in \cite{DeWitt:1967uc}, following the derivation in \cite{BroutEnglert:1966} for scalar matter. We basically extend here DeWitt's derivation to the vierbein formulation of the theory.

Let us consider the self-energy of the matter fields $\psi_\alpha$ in the presence of a background vierbein and a background scalar, and a vertex with an additional vierbein line,
\begin{equation}\label{eq:pi}
	\Pi^{\beta\alpha}(y,x)\equiv \frac{\delta^2 \Gamma}{\delta \bar{\psi}_\alpha(x) \delta\bar{\psi}^\dag_\beta(y)},
	\quad 
	(\Gamma^{\beta\alpha}_e)^\mu{}_a(z;y,x)
	\equiv
	\frac{1}{2M_P}\frac{\delta \Pi^{\beta\alpha}(y,x)}{\delta e_\mu{}^a(z)}.
\end{equation}
Because the effective action is invariant under diffeomorphisms it does not change under the infinitesimal transformation (\ref{eq:diffs}), 
\begin{equation}
	\int d^dz \left[
	\frac{\delta\Gamma}{\delta\bar e_\mu{}^a(z)}\Delta\bar e_\mu{}^a(z)
	+\frac{\delta\Gamma}{\delta\bar\phi(z)}\Delta\bar\phi(z)
	+\Delta\bar\psi_\alpha(z)\frac{\delta\Gamma}{\delta\bar\psi_\alpha(z)}
	\right]=0.
\end{equation}
Therefore,  acting on this equation with two functional derivatives with respect to the matter fields  we obtain
\begin{equation}\label{eq:general identity}
	\int d^d z\, \left[
	\frac{\delta \Pi^{\beta\alpha}(y,x)}{\delta\bar{e}_\mu{}^a(z)}\Delta\bar{e}_\mu{}^a(z)
	+\frac{\delta \Pi^{\beta\alpha}(y,x)}{\delta\bar{\phi}(z)}\Delta\bar{\phi}(z)
	+ \frac{\delta\Delta\bar\psi_\gamma(z)}{\delta\bar{\psi}_\alpha(x)}
	\Pi^{\beta\gamma}(y,z)
	+\frac{\delta\Delta\bar\psi^\dag_\gamma(z)}{\delta\bar{\psi}^\dag_\beta(y)}\Pi^{\gamma\alpha}(z,x) \right]=0.
\end{equation}
Using the transformation (\ref{eq:diffs}) and the definitions  (\ref{eq:pi}), and evaluating the last equation in our background (\ref{eq:background})  we then get
\begin{equation}\label{eq:Ward h}
	 2M_P\int d^d z \,\delta_\nu{}^a \xi^\nu(z) 
	\frac{\partial}{\partial z^\mu}(\Gamma^{\beta\alpha}_e)^\mu{}_a(z;y,x)
	+\frac{\partial}{\partial y^\mu}\left[\xi^\mu(y)  \Pi^{\beta\alpha}(y,x)\right]
	+\frac{\partial}{\partial x^\mu} \left[\xi^\mu(x)  \Pi^{\beta\alpha}(y,x)\right]=0.
\end{equation}
 In momentum space, with our  momentum conventions (\ref{eq:momentum conventions}), this becomes the identity
\begin{equation}
	2M_P (p'_\mu-p_\mu) (\Gamma_e^{\beta\alpha})^\mu{}_\nu (p',p)=
	p'_\nu \, \Pi^{\beta\alpha}(p)
	-p_\nu\, \Pi^{\beta\alpha}(p'),
\end{equation}
which in the limit of zero momentum transfer $p'\to p$ and after symmetrization reduces to the Ward-Takahashi identity for graviton emission,
\begin{equation}\label{eq:Ward q=0}
	2M_P(\Gamma_h^{\beta\alpha})^{\mu\nu} (p,p)=\Pi^{\beta\alpha}(p)\, 
	\eta^{\mu\nu} 
	-p^{(\mu} \frac{\partial \Pi^{\beta\alpha}}{\partial p_{\nu)}}.
\end{equation}
An analogous identity holds in electromagnetism.  

The self-energy is the sum of the tree-level contribution $\pi^{\beta\alpha}$ and the sum of all one-particle-irreducible self-energy diagrams\footnote{For a scalar, $\pi\equiv - (2\pi)^d(p^2+m^2)$, and $\Delta \pi=(2\pi)^d \pi^*$, where $\pi^*$ is what is usually called the self-energy insertion \cite{Weinberg:1995mt}.}  $\Delta\pi^{\beta\alpha}$,
\begin{equation}\label{eq:pi decomposition}
	\Pi^{\beta\alpha}=\pi^{\beta\alpha}+\Delta\pi^{\beta\alpha}.
\end{equation}
It is convenient to work in renormalized perturbation theory, with fields whose self-energy corrections vanish on shell, and whose propagators have unit residue at the corresponding pole, 
\begin{equation}\label{eq:OS}
	\Delta\pi^{\beta\alpha}\Bigg|_{OS}=0, \quad
	\frac{\partial\Delta\pi^{\beta\alpha}}{\partial p_\mu} \Bigg|_{OS}=0.
\end{equation}
The irreducible vertex $(\Gamma^{\beta\alpha}_h)^{\mu\nu}$ is also  the sum of the tree contribution $(\gamma^{\beta\alpha}_h)^{\mu\nu}$ and the contribution from loop diagrams $(\Delta\gamma^{\beta\alpha}_h)^{\mu\nu}$,
\begin{equation}\label{eq:Gamma decomposition}
	(\Gamma^{\beta\alpha}_h)^{\mu\nu}=
	(\gamma^{\beta\alpha}_h)^{\mu\nu}+(\Delta\gamma^{\beta\alpha}_h)^{\mu\nu}.
\end{equation}
Because the Ward identity (\ref{eq:Ward q=0}) is merely an expression of diffeomorphism invariance, it also holds in the limit in which all coupling constants of the theory go to zero, in which  we can approximate all quantum amplitudes by tree-level expressions. Hence, the tree vertex and the tree-level self-energy  obey the identity (\ref{eq:Ward tree q=0}), the tree-level counterpart of equation (\ref{eq:Ward q=0}), as the reader can explicitly check.

We are ready now to derive the main result of this subsection. Substituting equations (\ref{eq:pi decomposition}) and (\ref{eq:Gamma decomposition}) into the Ward-Takahashi identity  (\ref{eq:Ward q=0}), using the tree-level relation (\ref{eq:Ward tree q=0}), and going on shell, equations (\ref{eq:OS}), we conclude that 
\begin{equation}\label{eq:h no corrections}
	(\Delta\gamma^{\beta\alpha}_h)^{\mu\nu}\Bigg|_{OS}=0. 
\end{equation}
On shell, and in the limit of zero-momentum transfer, quantum corrections to the gravitational vertex vanish.  Since, as we have seen in Subsection \ref{sec:The Weak Equivalence Principle}, tree-level (classical) amplitudes do respect the equivalence principle, so do the quantum corrected ones. As before, this result has an analogous counterpart in electromagnetism, which guarantees  the non-renormalization of the  electric charge (up to an overall wave function renormalization constant) at zero momentum transfer.

\subsection{Scalar Emission}
\label{sec:Scalar Emission}

Let us turn our attention now to the emission of a scalar  by matter. Although there is no analogous Ward identity for scalar emission, because of  the structure of the couplings of $\phi$ to matter the vertex for scalar emission is closely related to that for graviton emission, whose properties it partially inherits. To see this, note that the matter action $S_M^E$ in (\ref{eq:S EF M}) is  invariant under the set of infinitesimal transformations
\begin{subequations}\label{eq:redef}
\begin{align}
	&\psi_\alpha\to \psi_\alpha'= \psi_\alpha, \\
	& \phi\to \phi'=\phi+ \epsilon \,  M, \\
	&e_\mu{}^a\to  e'_\mu{}^a=
	e_\mu{}^a-\epsilon \, \frac{f'}{f} e_\mu{}^a,
	\label{eq:e redef}
\end{align}
\end{subequations}
where $\epsilon$ is an arbitrary function on spacetime.  For certain functions $f(\phi/M)$, namely, exponentials, this transformation can be promoted to a group of $U(1)$ transformations that act on $\phi$ by a shift, and on the vierbein by  a Weyl transformation. In that particular case, the transformations (\ref{eq:redef}) are linear in the fields, though, in general, the transformation (\ref{eq:e redef}) is non-linearly realized.  Whatever the case, if (\ref{eq:redef}) were an exact, linearly-realized, global symmetry of the full action we would get, plugging the transformation rules (\ref{eq:redef}) into the general identity (\ref{eq:general identity}), and evaluating at our background (\ref{eq:background})
\begin{equation}\label{eq:approximate}
\int d^dz\left[	M\frac{\delta\Pi^{\beta\alpha}(y,x)}{\delta\bar{\phi}(z)}
	-\frac{f'(0)}{f(0)}
	\frac{\delta\Pi^{\beta\alpha}(y,x)}{\delta \bar{e}_\mu{}^a (z)}
	\delta_\mu{}^a\right]=0,
\end{equation}
where we have used that linear symmetries of the action are symmetries of the effective action.
Using equations (\ref{eq:vertices}) and (\ref{eq:self-energies}) this would lead immediately to the zero momentum  identity
\begin{equation}\label{eq:correlation}
	M\, \Gamma^{\beta\alpha}_\phi(p,p)=
	2M_P (\Gamma^{\beta\alpha}_h)^\mu{}_\mu(p,p),
\end{equation}
which relates the vertex for scalar emission to that for graviton emission. Since the latter satisfies equation (\ref{eq:Ward q=0}) it would then follow  in the limit of zero momentum transfer that
\begin{equation}
	M\, \Gamma^{\beta\alpha}_\phi(p,p)
	=d\, \Pi^{\beta\alpha}(p)
	-p^{\mu} \frac{\partial \Pi^{\beta\alpha}}{\partial p^{\mu}},
\end{equation}
and, as in the graviton case, using the tree-level relation (\ref{eq:Ward phi tree}) this would finally yield
	\begin{equation}
\Delta \gamma_\phi\Bigg|_{OS}=0,
\end{equation}
which states that quantum corrections to the scalar vertex in the limit of zero-momentum transfer vanish. Since the tree-level scalar vertex does respect the equivalence principle, so would quantum corrections. Note that if, in addition, the transformation (\ref{eq:redef}) were a local symmetry, we would be able to eliminate $\phi$ altogether from the theory by choosing the appropriate gauge. 

But, of course, the full action is not invariant under the global transformation (\ref{eq:redef}), and moreover, in general, the transformation (\ref{eq:redef}) is non-linear. From this point of view,  equation (\ref{eq:approximate}) is just an approximation to zeroth order in symmetry-breaking terms of a  general Ward-Takahashi identity that  we derive in Appendix \ref{sec:Ward identities for broken symmetries}. To apply the general Ward identity (\ref{eq:WI}) to our case, consider a linear version of the Weyl transformation (\ref{eq:redef}) acting on the field fluctuations,\begin{subequations}\label{eq:redef linear}
\begin{align}
	&\phi\to \phi'=\phi+ \epsilon\, M,\\
	&\delta e_\mu{}^a\to \delta e'_\mu{}^a= \delta e_\mu{}^a
	-\epsilon\, \frac{f'(0)}{f(0)}(\bar{e}_\mu{}^a+\delta e_\mu{}^a) .
\end{align}
\end{subequations}
Using the normalization conditions (\ref{eq:f normalization}), substituting equations (\ref{eq:redef linear}) into (\ref{eq:WI}), and taking two functional derivatives with respect to the matter fields yields the analogue of equation (\ref{eq:correlation}), modulo corrections due to the fact that the transformation (\ref{eq:redef}) is not an exact (linear) symmetry of the action,
\begin{equation}\label{eq:exact}
	M \, \Gamma^{\beta\alpha}_\phi(p,p)=
	2M_P (\Gamma^{\beta\alpha}_h)^\mu{}_\mu(p,p)+
	\Gamma^{\beta\alpha}_{\Delta}.
\end{equation}
Here, as we detail in Appendix \ref{sec:Ward identities for broken symmetries},  $\Gamma^{\alpha\beta}_\Delta$ is the sum of all one-particle-irreducible diagrams  with two external fields $\psi_\alpha$ and $\psi_\beta$ (with amputated propagators), and a vertex insertion of $\Delta$, the change of the Lagrangian density under the linear transformation (\ref{eq:redef linear}), carrying zero momentum into the diagram.

In order to determine the explicit form of $\Delta$  we note that, from the action (\ref{eq:S EF}),
\begin{subequations}
\begin{eqnarray}\label{eq:R contr}
	\frac{\delta S_\phi}{\delta\phi}&=& \det e \, 
	\left[\Box \phi-\frac{dV}{d\phi}\right],
	\label{eq:Sphiphi}\\
	e_\mu{}^a \frac{\delta S_\phi}{\delta e_\mu{}^a}
	&=&-\det e \left[\frac{d-2}{2}\partial_\mu\phi \partial^\mu \phi
	+ d \, V(\phi)\right],
	\label{eq:Sphie}\\
	e_\mu{}^a \frac{\delta S_{EH}}{\delta\, \delta  e_\mu{}^a(z)}&=&
	\det e\, \frac{(d-2)\,M_P^2}{2}  R,
	\label{eq:SEH}\\
	e_\mu{}^a \frac{\delta S_M}{\delta\, \delta  e_\mu{}^a}&\equiv& 
	\det e \, f^d \, T_M{}^\mu{}_\mu,
	\label{eq:SM}\\
	\frac{\delta S_M}{\delta\phi}&=& 
	\det e \, f^d \frac{f'}{M f}\, T_M{}^\mu{}_\mu, 
\end{eqnarray}
\end{subequations}
where $R$ the scalar curvature and 
\begin{equation}
	(T_M)_\mu{}^\nu \equiv \frac{f\, e_\mu{}^a }{\det (f\, e)}
	\frac{\delta S_M}{\delta (f e_\nu{}^a)} 
\end{equation}	
is the energy-momentum tensor of matter, which depends on $\phi$ because we assume that the matter action is of the form (\ref{eq:S EF M}).  Hence, using equations (\ref{eq:delta def}), (\ref{eq:redef linear}) and (\ref{eq:f normalization}) we arrive at
\begin{equation}\label{eq:delta S tot conformal}
	\Delta =
	M\frac{\delta S_\phi}{\delta\phi(x)}-e_\mu{}^a
	\frac{\delta (S_\phi+\delta S_{EH}+\delta S_{GF})}
	{\delta\, \delta e_\mu{}^a(x)}
	+\det e \, f^d \left(\frac{f'}{f}-1\right)\, T_M{}^\mu{}_\mu,
\end{equation}
which we should expand around our background (\ref{eq:background}) in order to calculate the corresponding diagrams.  The key of this result is that the correction term proportional to the energy-momentum tensor of matter is at least proportional to $\phi$. This reflects that the transformation (\ref{eq:redef linear}) leaves the part of the matter action proportional to $f(0)$ invariant. 

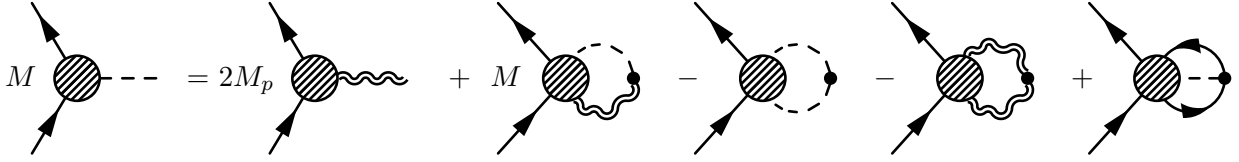
\begin{figure}
$M$  \!\!\!\!\!\!\!\!\!\! \subfigure
{
\parbox{20mm}{\begin{fmfgraph*}(20,20) 
\fmfleft{i1,i2} \fmfright{o1}
\fmfblob{.3w}{v}
\fmf{fermion}{i1,v,i2}
\fmf{dashes}{v,o1}
\end{fmfgraph*}}
}
=
$2 M_p$ \!\!\!\!\!\!\!\!\!\! \subfigure
{
\parbox{20mm}{\begin{fmfgraph*}(20,20) 
\fmfleft{i1,i2} \fmfright{o1}
\fmfblob{.3w}{v}
\fmf{fermion}{i1,v,i2}
\fmf{dbl_wiggly}{v,o1}
\end{fmfgraph*}}
}
\,\,\, $+ \,\,\,\, M$ \!\!\!\!\!\!\!\!\!\!\!\!\!\! \subfigure
{
\parbox{20mm}{\begin{fmfgraph*}(20,20) 
\fmfleft{i1,i2} \fmfright{o1}
\fmfblob{.3w}{v}
\fmf{fermion}{i1,v,i2}
\fmf{dbl_wiggly,left}{o1,v}
\fmf{dashes,right}{o1,v}
\fmfdot{o1}
\end{fmfgraph*}}
}
\,\,\, $-$ \!\!\!\!\!\!\!\!\!\!
\subfigure
{
\parbox{20mm}{\begin{fmfgraph*}(20,20) 
\fmfleft{i1,i2} \fmfright{o1}
\fmfblob{.3w}{v}
\fmf{fermion}{i1,v,i2}
\fmf{dashes,left}{o1,v}
\fmf{dashes,right}{o1,v}
\fmfdot{o1}
\end{fmfgraph*}}
}
\,\,\, $-$ \!\!\!\!\!\!\!\!\!\!
\subfigure
{
\parbox{20mm}{\begin{fmfgraph*}(20,20) 
\fmfleft{i1,i2} \fmfright{o1}
\fmfblob{.3w}{v}
\fmf{fermion}{i1,v,i2}
\fmf{dbl_wiggly,left}{o1,v}
\fmf{dbl_wiggly,right}{o1,v}
\fmfdot{o1}
\end{fmfgraph*}}
}
\,\,\, $+$ \!\!\!\!\!\!\!\!\!\!
\subfigure
{
\parbox{20mm}{\begin{fmfgraph*}(20,20) 
\fmfleft{i1,i2} \fmfright{o1}
\fmfblob{.3w}{v}
\fmf{fermion}{i1,v,i2}
\fmf{fermion,right}{v,o1}
\fmf{dashes,tension=0}{o1,v}
\fmf{fermion,right}{o1,v}
\fmfdot{o1}
\end{fmfgraph*}}
}
\caption{Diagrammatic expression of equation (\ref{eq:exact}) to leading order in the gravitational couplings $M_P^{-1}$ and $M^{-1}$. The irreducible vertex for scalar emission  equals the trace of that for graviton emission plus or minus corrections terms. In each correction term, the blob  represents  the sum of all 1PI diagrams with the corresponding number of external lines and the vertex insertion marked by a dot. }
\label{fig:representation}
\end{figure}

A graphical representation of equation (\ref{eq:exact}) to leading order in the gravitational couplings is given in figure \ref{fig:representation}, and helps to understand the different corrections due the broken Weyl symmetry. In our background, $\delta S_\phi/\delta\phi$ contains a linear term in $\phi$, whose insertion in a vertex does not lead to any 1PI diagrams. The next contribution from $\delta S_\phi/\delta\phi$ stems from a term proportional to $\phi \, h /M_P$, and thus,  the sum of all diagrams with an insertion of $M\,\delta S_\phi/ \delta\phi$ and two external matter lines contributes a term of order $M_P^{-2}$ to the equation in figure \ref{fig:representation}. (The diagram only has two external matter lines, so the scalar and graviton lines must end at a vertex in the diagram. Since the latter respectively couple with strength $M^{-1}$ and $M_P^{-1}$, the suppression must be at least of order $M /M_P \times M^{-1}\times M_P^{-1} = M_P^{-2}$.) Similarly, because $e_\mu{}^a \delta S_\phi/\delta(\delta e_\mu{}^a)$ is at least  quadratic in the scalar $\phi$, insertion of this vertex yields a contribution of order $M^{-2}$, from  the vertices at which the two scalar lines must end.   In our background the variational derivative $e_\mu{}^a\delta S_{EH}/\delta(\delta e_\mu{}^a)$ is  at least quadratic in the graviton field, and, therefore, the vertex containing (\ref{eq:SEH}) yields a contribution of order $M_P^{-2}$, the same as that from the gauge fixing term,  which is also quadratic in the graviton. Since the ghost action does not contain $h_{\mu\nu}$ nor $\phi$, it is invariant under the transformation (\ref{eq:redef linear}).  Finally the vertex insertion proportional to $T_M^\mu{}_\mu$ in equation (\ref{eq:delta S tot conformal}) is linear in $\phi/M$, and thus contributes a correction of order $M^{-2}$ to the proper vertex, unless $f''(\bar\phi=0)=1$, for which this term would be proportional to $(\phi/M)^2$, and hence would contribute a factor of order $M^{-3}$. An extreme example of the latter is an exponential, for which the insertion  proportional to $T_M^\mu{}_\mu$ would be absent altogether. Overall,  because of (\ref{eq:Ward q=0}), this translates into the approximate scalar Ward-Takahashi  identity
\begin{equation}\label{eq:scalar Ward}
	M\, \Gamma_\phi^{\beta\alpha}(p,p)=
	d \, \Pi^{\beta\alpha}(p) 
	-p^\mu \frac{\partial\Pi^{\beta\alpha}}{\partial p^\mu}
	+\mathcal{O}(M_P{}^{-2})+\mathcal{O}(M^{-2}).
\end{equation}
Expanding the scalar vertex  on the left hand side of the last equation into a tree-level contribution $\gamma_\phi$ and loop corrections $\Delta\gamma_\phi$, using equations (\ref{eq:pi decomposition}) and (\ref{eq:OS}) for the right hand side, and employing that the tree-level couplings of the scalar do respect the equivalence principle, equation (\ref{eq:Ward phi tree}),  we thus finally get
\begin{equation}\label{eq:violation}
	\Delta \gamma_\phi\Bigg|_{OS}=
	\mathcal{O}(M^{-1} M_P{}^{-2})+\mathcal{O}(M^{-3}).
\end{equation}
Quantum corrections to scalar couplings do violate the equivalence principle, but by terms suppressed by three powers of the gravitational couplings. Since experimental constraints require $M_P\ll M$ \cite{Will:2005va}, the dominant violations are of order $M^{-1} M_P^{-2}$. Although we have assumed for concreteness that the dynamics of the  graviton and scalar fields  is described by equation (\ref{eq:S EH phi}), it is straightforward to extend our analysis to more general forms. As long as the latter do not preserve the Weyl symmetry (\ref{eq:redef linear}),  there should be violations of the weak equivalence principle in those theories too.

\subsection{Extension of the Weyl Symmetry to the Full Action}
\label{sec:Extension of the Conformal Symmetry to the Full Action}

We have previously noted that exponentials $f=\exp(\phi/M)$ play a special role in the action (\ref{eq:S EF M}), since for such functions the  Weyl transformation (\ref{eq:redef}) is a linearly realized, exact symmetry of the matter action, $\Delta S_M^E=0$. In this case, the last term in equation (\ref{eq:delta S tot conformal}) is absent, and the corresponding equivalence principle violating corrections to the scalar vertex proportional to $T_M^\mu{}_\mu$ vanish.  It is then natural to ask whether this Weyl symmetry can be extended to the rest of the action.

Consider first the scalar field action $S_\phi$. To render it invariant under the transformation (\ref{eq:redef}) we just need to interpret $\phi$ as the Goldstone boson of a spontaneously broken Weyl symmetry. In that case, a mass term is forbidden by the global shift symmetry $\phi\to \phi+\epsilon\, M$, and field derivatives need to enter with appropriate factors of $\exp(\phi/M)$,  
\begin{equation}
	\tilde{S}_\phi=-\frac{1}{2}\int d^d x \, \det e \, \exp\left[\frac{(d-2)\phi}{M}\right]
	g^{\mu\nu}\partial_\mu \phi \partial_\nu \phi.
\end{equation}
It is then easy to check then, that this new action is invariant under global Weyl transformations, $\Delta \tilde{S}_\phi=0$. In such a theory, the correction terms in equation (\ref{eq:delta S tot conformal}) coming from the change of $S_\phi$  under the Weyl transformation would  vanish. 

Along the same lines, we can also extend the Einstein-Hilbert action to a globally Weyl invariant expression,
\begin{equation}
	\tilde{S}_{EH}=\int d^d x \det e 
	\exp\left[\frac{(d-2)\phi}{M}\right] \frac{M_P^2}{2} R,
\end{equation} 
which, again remains invariant under (\ref{eq:redef linear}), $\Delta \tilde{S}_{EH}=0$. For such an action, the correction terms in (\ref{eq:delta S tot conformal}) stemming from the change of $\tilde{S}_{EH}$ would again vanish.

However, we cannot make the full action Weyl invariant while keeping intact its scalar-tensor nature. In fact, if the total action reads
\begin{equation}
 \tilde{S}_\mathrm{tot}=\tilde{S}_{EH}+\tilde{S_\phi}
 +S_M\left[\exp(\phi/M)e_\mu{}^a,\psi\right], 
\end{equation}
the field redefinition $\tilde{e}_\mu{}^a \equiv e^{\phi/M} e_\mu{}^a$ leads to
\begin{equation}
	\tilde{S}_\mathrm{tot}=
	\int d^d x \det \tilde{e}\left[\frac{M_P^2}{2}\tilde{R}
	-\frac{1}{2}\left(1+(d-2)(d-1)\frac{M_P^2}{M^2}\right)
	\tilde{g}^{\mu\nu}\partial_\mu \phi \partial_\nu \phi\right]
	+S_M[\tilde{e}_\mu{}^a, \psi].
\end{equation}
This is just the action of general relativity minimally coupled to matter with an extended matter sector consisting of a minimally coupled massless scalar.  Because there are no vertices with an odd power of $\phi$ in this theory, the amplitude for emission of a single scalar by matter vanishes (in any case, the scalar field couples derivatively, so it cannot mediate a long-ranged interaction.)

It is also instructive to consider the action (\ref{eq:S EF}) in flat space, with gravitation turned off ($M_P\to \infty$). Though the broken Weyl symmetry (\ref{eq:redef linear}) acts non-trivially on the metric, this approximate symmetry does not get lost. Indeed, with all matter fields taken to be  diffeomorphism scalars,  in flat spacetime and for an exponential $f$ the Weyl transformation (\ref{eq:e redef}) has the same effect on the vierbein as the infinitesimal coordinate dilatation 
\begin{equation}\label{eq:coordinate dilatation}
	x^\mu\to \left(1- \epsilon \, \frac{f'(0)}{f(0)}\right)x^\mu.
\end{equation}
Therefore, in that case, as a consequence of diffeomorphism invariance, the matter action in the Einstein frame possesses an exact dilatation symmetry under which the fields transform according to
\begin{subequations}
\begin{align}
	&\psi_\alpha\to \psi_\alpha+\epsilon\,  x^\mu\partial_\mu \psi_\alpha, \label{eq:dilatation unconv}\\
	&\phi\to \phi+\epsilon (M+ x^\mu\partial_\mu \phi),
\end{align}
\end{subequations}
where we have used the normalization conditions (\ref{eq:f normalization}).

The dilatation (\ref{eq:dilatation unconv}) does not act  conventionally on the matter fields. To bring it to its usual form it is convenient to redefine the matter fields. Suppose that the kinetic term of the matter field $\psi_\alpha$ contains $n$ derivatives. Then, diffeomorphism invariance implies that each derivative is accompanied by the inverse of the vierbein, and that the integration measure $d^d x$ is multiplied by $\det e$. Therefore, in the Einstein frame the kinetic term of the field $\psi_\alpha$ is proportional to $f^{d-n}$. Let us hence redefine
\begin{equation}\label{eq:matter field redef}
	\tilde{\psi}_\alpha=f^{\frac{d-n}{2}}\psi_\alpha.
\end{equation}
Then, by construction, the kinetic term of $\tilde{\psi}_\alpha$ does not contain factors of $f$ (though there may be additional derivative interactions), and the matter action is  invariant under
\begin{subequations}\label{eq:dilatation}
\begin{align}
	&\tilde{\psi}_\alpha\to \tilde{\psi}_\alpha
	+\epsilon\, \left(\frac{d-n}{2}
	+x^\mu\partial_\mu\right)
	\tilde{\psi}_\alpha,\\
	&\phi\to \phi+\epsilon(M+x^\mu\partial_\mu \phi),
\end{align}
\end{subequations}
where we have used again equation  (\ref{eq:f normalization}).
Acting on the matter fields, this is now  a conventional dilatation, since $(d-n)/2$ is the scaling dimension of the field $\psi_\alpha$. The inhomogeneous term in the transformation of $\phi$ underscores its interpretation as a pseudo Nambu-Goldstone boson of an approximate, spontaneously broken conformal symmetry, although, even for a massless $\phi$, the scalar field action is not invariant under (\ref{eq:dilatation}).

Because the field $\phi$ transforms inhomogeneously under (\ref{eq:dilatation}), the vertex for scalar emission satisfies a Ward-Takahashi identity (\ref{eq:WI}) related to this broken symmetry \cite{Coleman:1970je},
\begin{equation}\label{eq:dilatation WI}
	M \, \Gamma^{\beta\alpha}_\phi(p,p)
	+ \left(p^\mu \frac{\partial}{\partial p^\mu}-n\right)
	\Pi^{\beta\alpha}(p)=\Gamma^{\beta\alpha}_\Delta, 
\end{equation}
where, again, $\Gamma^{\alpha\beta}_\Delta$ is the sum of all 1PI diagrams with two external $\psi$ lines and a vertex insertion of $\Delta$, the change in the Lagrangian density under the infinitesimal transformation (\ref{eq:dilatation}). This equation is the flat space counterpart of equation (\ref{eq:scalar Ward}), and also guarantees that, for fields renormalized on shell, quantum corrections to the vertex for scalar emission are determined by the change of the action under the broken symmetry (\ref{eq:dilatation}). 

The dilatation (\ref{eq:coordinate dilatation}) is part of the conformal group, the set of all coordinate transformations that preserve the Minkowski metric up to an overall conformal factor. Along the same lines as for dilatations, as a consequence of diffeomorphism and Lorentz invariance, it is  easy to show that, for an exponential $f$, the matter action in flat space is symmetric under the full conformal group, acting again on the scalar $\phi$ linearly, but inhomogeneously. There exist then additional Ward identities related to the full conformal symmetry of the theory, though we shall not write them down. Although  conformal symmetries are typically anomalous, the structure of the couplings to $\phi$ in the matter action for an exponential $f$ guarantees that the  symmetry remains intact in the dimensionally regularized theory. If $f$ is not an exponential, or if scalar kinetic term is not conformally invariant,  conformal symmetry is broken, and the corresponding Ward identities contain the appropriate vertex insertions, as in the dilatation case.

\section{Specific Examples}

Our next goal is to illustrate our main results with a set of concrete examples that show the nature and size of the equivalence principle violations in scalar-tensor theories.  We only address this issue for scalars and spin half fermions; the vertex for scalar emission by a gauge boson vanishes on shell as a consequence of Lorentz and gauge invariance, so there is no need to consider this case in the context of the weak equivalence principle. 

We  first check explicitly in a one-loop calculation that couplings to matter do not lead to any violations of the weak equivalence principle. For scalars,  these results partially overlap and complement previous work in the literature \cite{Hui:2010dn}.  In addition, we verify that one-loop corrections involving the scalar $\phi$ do result in violations of the equivalence principle, in agreement with the Ward identity (\ref{eq:violation}).  To avoid the complications of index algebra, we focus on  loops of $\phi$ for simplicity, but we expect analogous violations from diagrams in which the loop contains at least one graviton.

\subsection{Scalar Matter}

Let us assume for the time being that matter consists of scalar particles $\chi$, which for simplicity interact through a cubic coupling with another species of scalar particles $\sigma$.   Then, in the Jordan frame, the matter action is
\begin{equation}
	\mathcal{L}_M^J=-\frac{1}{2}g^{\mu\nu}\partial_\mu\chi \partial_\nu \chi-\frac{1}{2}m^2 \chi^2-\frac{1}{2}g^{\mu\nu}\partial_\mu\sigma \partial_\nu \sigma-\frac{1}{2}m_\sigma^2 \sigma^2-\frac{\lambda}{2}\sigma \chi^2.
\end{equation} 
We are going to calculate quantum corrections to the vertex for emission of a scalar $\phi$ by matter $\chi$. In order to obtain the action in the Einstein frame, we apply  the conformal transformation implicit in (\ref{eq:S EF M}). As we have seen, exponentials play a somewhat special role in scalar-tensor theories, so, for the purposes of illustration we choose
\begin{equation}\label{eq:f exp}
	f\left(\frac{\phi}{M}\right)=\exp\left(\frac{\phi}{M}\right).
\end{equation}
Since we are interested in corrections to the vertex at most of order $1/M^3$ we then  expand the Einstein-frame action in Minkowski space  to third  order in $\phi$, and drop some of the terms that do not enter our calculation,
\begin{multline}\label{eq:scalar matter EF}
	\mathcal{L}_M^E=
	-\frac{1}{2}\partial_\mu \chi \partial^\mu \chi-\frac{m^2 }{2}\chi^2-
	\frac{1}{2}\partial_\mu \sigma \partial^\mu \sigma
	-\frac{m_\sigma^2}{2}\sigma^2-\frac{\lambda}{2}\sigma \chi^2
	-\frac{d\, \lambda}{2M}\phi \sigma \chi^2-\\
	-\frac{1}{2}\frac{\phi}{M}
	\left[(d-2)\partial_\mu\chi\partial^\mu\chi+d\,  m^2\chi^2
	\right]
	-\frac{1}{2}\frac{\phi}{M}
	\left[(d-2)\partial_\mu\sigma\partial^\mu\sigma+d\, m_\sigma^2\sigma^2
	\right]-\\
	-\frac{1}{4}\frac{\phi^2}{M^2}
	\left[(d-2)^2\partial_\mu \chi\partial^\mu\chi+d^2m^2 \chi^2\right]
	-\frac{1}{12}\frac{\phi^3}{M^3}
	\left[(d-2)^3\partial_\mu \chi\partial^\mu\chi+d^3 m^2 \chi^2\right]+\cdots.
\end{multline} 
Note that some of the couplings above are redundant, and can be removed away by a field redefinition. Although the field redefinition simplifies the Feynman rules, it somewhat obscures the symmetry between the couplings of the scalar and the graviton, so we shall mostly proceed with the Lagrangian  (\ref{eq:scalar matter EF}). Of course either formulations yield the same $S$-matrix elements. 

\subsubsection{Matter Loops}

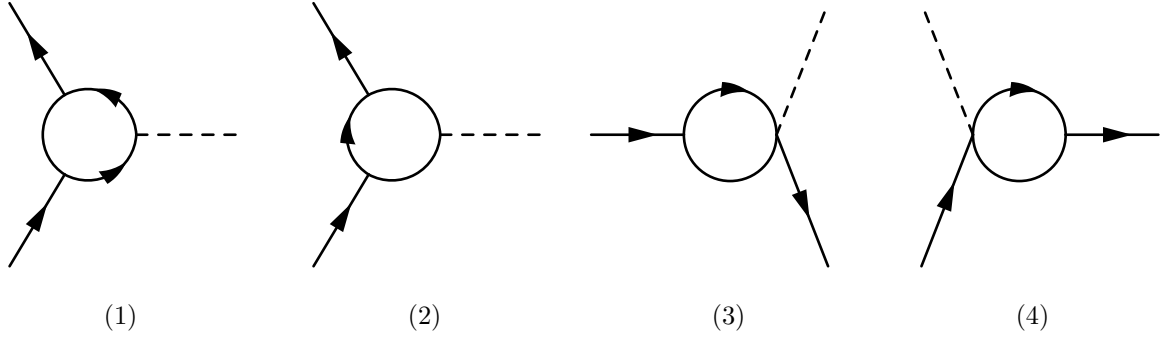
\begin{figure}
\renewcommand{\thesubfigure}{(\arabic{subfigure})}
\subfigure[]
{
\begin{fmfgraph}(35,35) 
\fmfleft{i1,i2} 
\fmfright{o1}
\fmf{fermion,tension=1.3}{i1,v1}
\fmf{fermion,tension=1.3}{v2,i2}
\fmf{vanilla,left=.55}{v1,v2}
\fmf{fermion,right=.55}{v1,v3,v2}
\fmf{dashes,tension=1.3}{v3,o1}
\end{fmfgraph}
}
\subfigure[]
{
\begin{fmfgraph}(35,35) 
\fmfleft{i1,i2} 
\fmfright{o1}
\fmf{fermion,tension=1.3}{i1,v1}
\fmf{fermion,tension=1.3}{v2,i2}
\fmf{fermion,left=.55}{v1,v2}
\fmf{vanilla,left=.55}{v2,v3,v1}
\fmf{dashes,tension=1.3}{v3,o1}
\end{fmfgraph}
}
\subfigure[]
{
\begin{fmfgraph}(35,35) 
\fmfleft{i1} 
\fmfright{o1,o2}
\fmf{fermion,tension=1.8}{v2,o1}
\fmf{fermion,tension=2}{i1,v1}
\fmf{fermion,left}{v1,v2}
\fmf{vanilla,left}{v2,v1}
\fmf{dashes,tension=1.8}{v2,o2}
\end{fmfgraph}
}
\subfigure[]
{
\begin{fmfgraph}(35,35) 
\fmfleft{i1,i2} 
\fmfright{o1}
\fmf{fermion,tension=1.8}{i1,v1}
\fmf{fermion,tension=2}{v2,o1}
\fmf{fermion,left}{v1,v2}
\fmf{vanilla,left}{v2,v1}
\fmf{dashes,tension=1.8}{v1,i2}
\end{fmfgraph}
}
\caption{One-loop corrections to the vertex for scalar emission by matter}
\label{fig:mcsv}
\end{figure}

Our first goal is to explicitly show that one-loop corrections in which matter fields run inside the loop do respect the equivalence principle. In order to do so, it is simpler (and more revealing) to verify first the Ward-Takahashi identity (\ref{eq:scalar Ward}).  Consider for that purpose the order $\lambda^2$ correction to the amplitude for emission of a scalar $\phi$ by a matter field $\chi$. At this order, the correction is given by the four diagrams in figure \ref{fig:mcsv}, where $\chi$ lines are labeled with an arrow, $\sigma$ lines are plain and $\phi$ lines are dashed. Because we are interested in the limit of zero momentum transfer, we consider equal  incoming and outgoing momenta. Using the vertices implied by the Lagrangian (\ref{eq:scalar matter EF}), and combining denominators using Feynman parameters in the standard way \cite{Weinberg:1995mt}, we find
\begin{subequations}\label{eq:scalar matter vertex corr}
\begin{eqnarray}
	i\Delta\gamma_1&=&-\frac{2\lambda^2}{M}\int d^d k \int_0^1 dx\, 
	\frac{x[(d-2)(p^2(1-x)^2+k^2)+d m^2]}{[k^2+p^2x(1-x)+m^2 x +m_\sigma^2(1-x)]^3},\\
	i\Delta\gamma_2&=& - \frac{2\lambda^2}{M}\int d^d k \int_0^1 dx\, 
	\frac{(1-x)[(d-2)(p^2 x^2+k^2)+d\, m_\sigma^2]}{[k^2+p^2x(1-x)+m^2 x +m_\sigma^2(1-x)]^3},\\
	i\Delta\gamma_3+i\Delta\gamma_4&=& \frac{2 d\,\lambda^2}{M}\int d^d k \int_0^1 dx\,
	\frac{1}{[k^2+p^2x(1-x)+m^2 x +m_\sigma^2(1-x)]^2},
\end{eqnarray}
\end{subequations}
where we have dropped the $i\epsilon$ factors in the propagators. Combining all the contributions in (\ref{eq:scalar matter vertex corr}) we thus conclude that the total vertex correction  is 
\begin{equation}\label{eq:scalar matter tot vertex}
	i\Delta\gamma\equiv i\sum_{i=1}^4 \Delta\gamma_i
	=\frac{\lambda^2}{M}\int d^d k \int_0^1 dx\,
	\frac{4k^2+4p^2x(1-x)}{[k^2+p^2x(1-x)+m^2 x +m_\sigma^2(1-x)]^3}.
\end{equation}

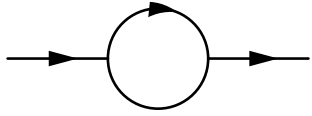
\begin{figure}
\begin{fmfgraph}(40,15) 
\fmfleft{i1}
\fmfright{o1}
\fmf{fermion,tension=2}{i1,v1}
\fmf{fermion,tension=2}{v2,o1}
\fmf{fermion,left}{v1,v2}
\fmf{vanilla,left}{v2,v1}
\end{fmfgraph}
\caption{One loop correction to the self-energy of matter}
\label{fig:mse}
\end{figure}

The interactions of matter $\chi$ with the field $\sigma$ also modify the self-energy of matter. At order $\lambda^2$, the self-energy corrections  are described by the diagram in figure \ref{fig:mse}, which leads to
\begin{subequations}\label{eq:scalar matter all}
\begin{equation}\label{eq:scalar matter self}
	i\Delta\pi=\lambda^2 \int d^d k \int_0^1 dx\,
	\frac{1}{[k^2+p^2x(1-x)+m^2 x+m_\sigma^2 (1-x)]^2},
\end{equation}
and directly yields
\begin{multline}\label{eq:scalar matter pdp}
	i\left(d \Delta\pi- p^\mu \frac{\partial \Delta\pi}{\partial p^\mu}\right)
	=\lambda^2 \int d^d k \int_0^1 dx\bigg(
	\frac{d}{[k^2+p^2x(1-x)+m^2 x +m_\sigma^2(1-x)]^2}
	+\\
	{}+\frac{4p^2x(1-x)}{[k^2+p^2x(1-x)+m^2 x +m_\sigma^2(1-x)]^3}\bigg).
\end{multline}
\end{subequations}

The integrals over loop momenta in equations (\ref{eq:scalar matter tot vertex}) and (\ref{eq:scalar matter all})  can be explicitly carried out by rotating the integration contour counterclockwise into Euclidean momenta and making use of the well known relation
\begin{equation}\label{eq:momentum int}
	\int d^d k_E \frac{(k^2)^n}{[k^2+\Delta^2]^m}=\pi^{d/2}
	\frac{\Gamma(\frac{d+2n}{2})}{\Gamma(\frac{d}{2})}
	\frac{\Gamma(m-\frac{d+2n}{2})}{\Gamma(m)}\Delta^{d+2n-2m},
\end{equation}
which immediately confirms the Ward identity (\ref{eq:scalar Ward}).

When we calculate $S$-matrix elements (as opposed to Green's functions) it is convenient to work in the OS scheme of renormalized perturbation theory. We then need to introduce appropriate field renormalization and mass counterterms to enforce our renormalization conditions (\ref{eq:OS}). With $\chi\to  Z^{1/2} \chi$ and $m^2\to m^2-\delta m^2$  the counterterm Lagrangian becomes
\begin{eqnarray}\label{eq:scalar C}
	\mathcal{L}_C^E=&-&\frac{1}{2}(Z-1)(\partial_\mu \chi \partial^\mu \chi +m^2 \chi^2)
	+\frac{1}{2}Z\delta m^2\chi^2- \nonumber\\
	&-&\frac{1}{2}\frac{\phi}{M}\left\{(Z-1)[(d-2)\partial_\mu\chi\partial^\mu \chi+ d \, m^2 \chi^2]
	-d\, Z \delta m^2 \chi^2\right\}+\cdots,
\end{eqnarray}
with $Z$ and $\delta m^2$ chosen to satisfy the conditions (\ref{eq:OS}),
\begin{equation}\label{eq:scalar CT}
	Z-1=\frac{1}{(2\pi)^d}\frac{d\Delta\pi}{dp^2}\Bigg|_{p^2=-m^2}, \quad 
	Z\delta m^2=-\frac{\Delta \pi(-m^2)}{(2\pi)^d}.
\end{equation}
These counterterms yield the additional contributions to the vertex amplitude
\begin{equation}\label{eq:delta gamma 5}
	i\Delta\gamma_5=-i \frac{(2\pi)^d}{M}\left\{(Z-1)
	\left[(d-2)p^2+d\, m^2\right]- d\, Z\delta m^2\right\}.
\end{equation}

Using the Ward identity (\ref{eq:scalar Ward}), evaluated at $p^2=-m^2$, it is now straightforward to see that the total vertex correction vanishes.  Alternatively,  bringing all the factors in $\Delta\gamma_i$ to a common denominator, and simplifying the resulting numerator we find that the total vertex correction is
\begin{equation}\label{eq:scalar matter delta vertex}
	i(\Delta \gamma_\phi)_{OS} \equiv i\sum_{i=1}^5
	\Delta \gamma_i=\frac{\lambda^2}{M}
	\int d^d k \int_0^1 dx\,
	\frac{(4-d) k^2 - d [m^2 x^2+ m_\sigma^2 (1-x)]}{[k^2+m^2x^2+m_\sigma^2(1-x)]^3}.
\end{equation}
Using equation (\ref{eq:momentum int}) in (\ref{eq:scalar matter delta vertex})  yields again $(\Delta\gamma_\phi)_{OS}=0$, in agreement with our general result (\ref{eq:violation}). The corresponding cancellation among the five different diagrams is an expression of diffeomorphism and Weyl invariance. In the Lagrangian (\ref{eq:scalar matter EF}), the vertex to which a single scalar $\phi$ is attached could be replaced by one to which a single graviton is attached. Since the Ward identity (\ref{eq:h no corrections})  guarantees that the sum of all diagrams that contribute to  the vertex correction for graviton emission vanishes in the appropriate kinematic limit, this result  transfers to the vertex for emission of a scalar particle.

This also explains why the total vertex correction does not vanish if we simply use a cut-off to regularize the theory.  If we cut off the Euclidean momentum integrals at $k_E=\Lambda$ in $d=4$  we get, from equation (\ref{eq:scalar matter delta vertex}),
\begin{equation}\label{eq:cutoff violation}
	(\Delta \gamma_\phi)_{OS}= -\frac{2\pi^2\lambda^2}{M} \int_0^1 dx\,
\left(1+\frac{m^2 x^2+m_\sigma^2 (1-x)}{\Lambda^2}\right)^{-2}.
\end{equation}
This remains finite in the limit $\Lambda\to \infty$, but does not vanish. The origin of the non-zero correction is of course the breaking of diffeomorphism invariance by the momentum cut-off, which leads to a breakdown of the Ward-Takahashi identity for graviton emission (\ref{eq:Ward tree q=0}), but does not affect the relation (\ref{eq:phi h}) between the vertex and the graviton vertex.  Although the quantum theory of massless spin two particles with non-derivative couplings to matter requires diffeomorphism invariance \cite{Weinberg:1965rz}, the coupling of a spin zero scalar $\phi$ to matter does not demand any symmetry. In other words, by regulating the momentum integrals with a cut-off, we are not breaking any symmetry in the scalar sector that is not already broken, so a momentum cut-off appears to be a perfectly valid regularization method. In this light, even our claim that matter loops do respect the equivalence is somewhat misleading.  

\subsubsection{Scalar Loops}
\label{sec:scalar loops}

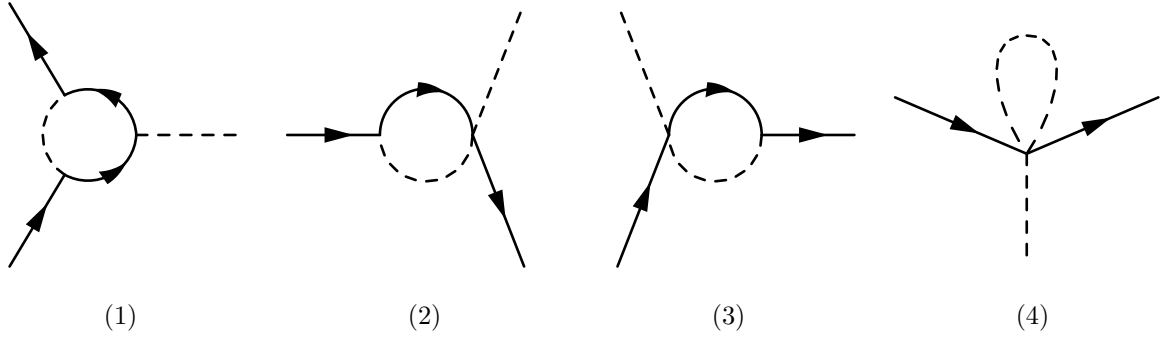
\begin{figure}
\subfigure[]
{
\begin{fmfgraph}(35,35) 
\fmfleft{i1,i2} 
\fmfright{o1}
\fmf{fermion,tension=1.3}{i1,v1}
\fmf{fermion,tension=1.3}{v2,i2}
\fmf{dashes,left=.55}{v1,v2}
\fmf{fermion,right=.55}{v1,v3,v2}
\fmf{dashes,tension=1.3}{v3,o1}
\end{fmfgraph}
}
\subfigure[]
{
\begin{fmfgraph}(35,35) 
\fmfleft{i1} 
\fmfright{o1,o2}
\fmf{fermion,tension=1.8}{v2,o1}
\fmf{fermion,tension=2}{i1,v1}
\fmf{fermion,left}{v1,v2}
\fmf{dashes,left}{v2,v1}
\fmf{dashes,tension=1.8}{v2,o2}
\end{fmfgraph}
}
\subfigure[]
{
\begin{fmfgraph}(35,35) 
\fmfleft{i1,i2} 
\fmfright{o1}
\fmf{fermion,tension=1.8}{i1,v1}
\fmf{fermion,tension=2}{v2,o1}
\fmf{fermion,left}{v1,v2}
\fmf{dashes,left}{v2,v1}
\fmf{dashes,tension=1.8}{v1,i2}
\end{fmfgraph}
}
\subfigure[]
{
\begin{fmfgraph}(35,45) 
\fmfleft{i1} \fmfright{o1} \fmfbottom{s} 
\fmf{fermion}{i1,vs,o1}
\fmf{dashes,tension=.75}{vs,vs}
\fmf{dashes}{vs,s}
\end{fmfgraph}
}
\caption{One-loop corrections to the vertex for scalar emission at order $1/M^3$. Continuous lines denote matter fields, while dashed lines label the scalar $\phi$. }
\label{fig:scsv}
\end{figure}

We proceed now to calculate  corrections to the vertex that include the scalar $\phi$ running inside a loop. These are described by the four diagrams in figure \ref{fig:scsv}, which respectively lead to the four vertex corrections
\begin{subequations}\label{eq:scalar delta vertex}
\begin{align}
	&i\Delta\gamma_1=-\frac{2}{M^3}\int d^d k \, dx\, x
	\frac{[(d-2)p\cdot(p(1-x)-k)+d\,  m^2]^2
	[(d-2)(p(1-x)-k)^2+d\, m^2]}
	{[k^2+p^2 x(1-x)+m^2 x+m_\phi^2 (1-x)]^3},
	\label{eq:scalar delta vertex 1}\\
	&i\Delta\gamma_2=\frac{1}{M^3}\int d^d k \, dx\, 
	\frac{\left[(d-2)p(p(1-x)-k)+d\, m^2\right]
	\left[(d-2)^2 \,p(p(1-x)-k)+d^2\,m^2\right]}
	{[k^2+p^2x(1-x)+m^2 x+m_\phi^2(1-x)]^2},\\
	&i\Delta\gamma_3=i\Delta\gamma_2,\\
	&i\Delta\gamma_4=-\frac{1}{2M^3}
	\int d^d k\, \frac{(d-2)^3p^2+d^3\, m^2}{k^2+m_\phi^2},
\end{align}
\end{subequations}
where, from now on and as before, the integral over $x$ covers the range from zero to one. 

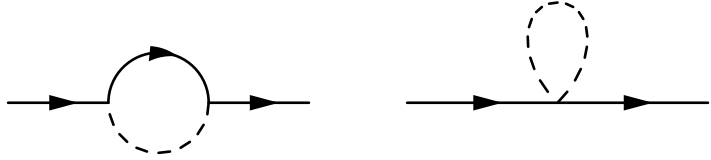
\begin{figure}
\subfigure
{
\begin{fmfgraph}(40,15) 
\fmfleft{i1}
\fmfright{o1}
\fmf{fermion,tension=2}{i1,v1}
\fmf{fermion,tension=2}{v2,o1}
\fmf{fermion,left}{v1,v2}
\fmf{dashes,left}{v2,v1}
\end{fmfgraph}
}
\qquad
\subfigure
{
\begin{fmfgraph}(40,15) 
\fmfleft{i1} \fmfright{o1}
\fmf{fermion}{i1,v,o1} 
\fmf{dashes}{v,v}
\end{fmfgraph}
}
\caption{Self-energy of  matter to order $1/M^2$.}
\label{fig:sse}
\end{figure}

Because we want to show that $\phi$ loops do lead to violations of the equivalence principle, it is more convenient to work in an on-shell renormalization scheme (OS).  The self-energy insertion $\Delta\pi$ is determined  by the two diagrams in figure \ref{fig:sse}, and the corresponding corrections read
\begin{align}\label{eq:scalar se}
	&i\Delta\pi_1=\frac{1}{M^2}\int d^d k \int_0^1 dx\,
\frac{[(d-2)p\cdot(p(1-x)-k)+d\,m^2]^2}{[k^2+p^2 x(1-x)+m^2 x+m_\phi^2 (1-x)]^2},\\
	&i\Delta\pi_2=-\frac{1}{2M^2}\int d^d k \, \frac{(d-2)^2 p^2+d^2 m^2}{k^2+m_\phi^2}.
\end{align}
In order to enforce the renormalization conditions (\ref{eq:OS}), we introduce a renormalized field $\chi\to Z^{1/2}\chi$ and a renormalized mass $m^2\to m^2-\delta m^2$, which give the counterterms in the Lagrangian  (\ref{eq:scalar C}). But  because we are dealing now with  non-renormalizable interactions (operators of mass dimension higher than $d$), the self-energy also contains a divergent term proportional to $p^4$, which we cannot absorb simply by renormalization of  fields and parameters present in the action (\ref{eq:scalar matter EF}). We are thus forced to introduce a new bare term with four derivatives and two fields, which we treat as a perturbation. In  the Jordan frame Lagrangian, this can be taken to be proportional to $\det e\, (\Box \chi)^2$, which in the Einstein frame becomes
\begin{equation}\label{eq:Z delta c}
	\mathcal{L}_C^E\supset \frac{Z \, \delta c}{2} \, f^{d-4}\cdot  (\Box \chi)^2,
\end{equation}
with  $Z\delta c$  chosen to enforce for instance the additional renormalization condition
\begin{equation}
	\frac{d\Delta\pi}{d(p^4)}\Bigg|_{p^2=-m^2}=0.
\end{equation}
(For simplicity we assume that the renormalized $c$ vanishes.) The  counterterms then yield additional vertex corrections, as in equation (\ref{eq:delta gamma 5}), but with the additional contribution from (\ref{eq:Z delta c})
\begin{equation}
	i\Delta\gamma_5=-i \frac{(2\pi)^d}{M}\left\{(Z-1)
	\left[(d-2)p^2+d\, m^2\right]- d\, Z\delta m^2
	-(d-4) Z\, \delta c \, p^4 \right\}.
\end{equation}
From the structure of the self-energy corrections, it is clear that the counterterms are of order $M^{-2}$.

We are ready to compute now the total correction to the vertex $(\Delta\gamma)_{OS}=\sum_i\Delta\gamma_i$. To make our point, let us concentrate of the phenomenologically relevant case of $d=4$ dimensions. In this limit, some of the momentum integrals diverge. It is relatively easy to isolate the residue of the pole as $d\to 4$, which, in the limit $m_\phi=0$ and after performing a trivial integral over $x$ reads
\begin{equation}\label{eq:pole scse}
	(\Delta\gamma_\phi)_{OS}=-\frac{4\pi^2}{M^3} \frac{16m^4+7m^2p^2+p^4}{d-4}
	+\mathcal{O}[(d-4)^0].
\end{equation}
The form of this pole immediately reveals that the theory defined by the action (\ref{eq:S EF M}) is non-renormalizable, in the broad sense that we cannot absorb its divergences  by appropriate renormalization of the coupling constants and parameters appearing in \emph{any} matter action of the form  (\ref{eq:S EF M}). Say, suppose that we introduce a renormalized coupling constant  by replacing $M^{-1}\to M^{-1}-\delta  M^{_-1}$. This introduces  additional counterterms in our theory, which to leading order in $1/M$ yield an additional vertex correction
\begin{equation}\label{eq:delta M}
 i\Delta\gamma_6=-i(2\pi)^d \delta M^{-1} 
 \left[(d-2)p^2 +d\,  m^2 \right].
\end{equation}
But comparison of equation (\ref{eq:pole scse}) with (\ref{eq:delta M}) quickly reveals that no single choice of $\delta M^{-1}$ cancels all the residues  at $d=4$, and that, in fact, we would have to choose three independent counterterms to cancel the terms proportional to $m^4$, $m^2 p^2$ and $p^4$. This means  that our theory contains three independent coupling constants, instead of one, as we initially thought.

What we are seeing here is that there is no symmetry that enforces the structure (\ref{eq:S EF M}) in scalar-tensor theories. In order to carry out the renormalization program we have to introduce all the terms compatible with the symmetries of the theory, which in this case \emph{only} consists of diffeomorphism invariance. In particular, just in the scalar sector alone, we have to introduce a set of coupling constant $1/M^{(j)}_i$ for each linear coupling of $\phi$ to an operator quadratic in the scalar matter species $\chi_i$,
\begin{equation}\label{eq:plethora}
	\mathcal{L}_M^E\to \sum_i \left[-\frac{1}{2}\partial_\mu \chi_i \partial^\mu \chi_i-\frac{1}{2}m_i^2 \chi_i^2
	-\frac{1}{2} \frac{\phi}{M^{(0)}_i} m_i^2\chi_i^2
	-\frac{1}{2}\frac{\phi}{M^{(2)}_i}\partial_\mu\chi_i\partial^\mu\chi_i
	-\frac{1}{2}\frac{\phi}{M^{(4)}_i}(\Box\chi_i)^2+\cdots
	\right].
\end{equation}
Because no common choice for all counterterms $\delta M_i^{(k)}$ can eliminate all the contributions to the pole at $d=4$ in equation (\ref{eq:pole scse}) for all matter species, and because the beta functions of the different coupling constants are determined by the coefficients of this pole \cite{'tHooft:1973mm}, these different couplings run differently with scale under the renormalization group flow.  Thus, once we include quantum corrections, the structure of (\ref{eq:S EF}) becomes untenable. 
The unnatural structure of the subclass  of scalar-tensor theories we consider here has been repeatedly emphasized by Damour (see e.g. \cite{Damour:2001fn}).

Let us proceed anyway  with the vertex correction and study its finite piece in the limit $d\to 4$. To simplify  the  algebra, we consider now on-shell momenta, $p^2=-m^2$ and focus on the limit $m_\phi=0$. In this case,  the finite terms reduce to
\begin{equation}\label{eq:OS violation}
	(\Delta\gamma_\phi)_{OS}=\mathcal{O}\left(\frac{1}{d-4}\right)-
	\frac{4\pi^2}{M^3}m^4\left[2+5\gamma+5\log (\pi m^2)\right],
\end{equation}
which again differs from zero. Of course, we should expect similar terms from the renormalization prescription that eliminates the pole at $d=4$.  Although we have explicitly calculated the corrections of order $(m/M)^3$, due to a scalar loop, we also expect non-vanishing corrections  of order $m^3/(M M_P^2)$ due to a graviton loop, as we argued in Section \ref{sec:Scalar Emission}. 

Equations (\ref{eq:pole scse}) and (\ref{eq:OS violation})  explicitly show that quantum corrections in scalar-tensor theories generically lead to violations of the equivalence principle. Of course, to make a precise and definite prediction about the size of these violations, we need to specify a renormalization prescription to eliminate the poles at $d=4$. In the absence of such a prescription, and on dimensional grounds, we generically expect the contribution of the scalar vertex to these violations to be of order $m^4/(M M_P^2)$ (to obtain the scattering amplitude one has to multiply this number by two powers of the appropriate mode function $u\propto 1/\sqrt{2p^0}\approx 1/\sqrt{2m}$). In that case, particles with different masses  fall with different accelerations.   In order to quantify the corresponding violations of the equivalence principle, it is conventional to quote the E\"otv\"os parameter $\eta$, defined to be the relative difference in acceleration of two different test bodies $A$ and $B$,
\begin{equation}
	\eta_{AB}=2\frac{a_A-a_B}{a_A+a_B}.
\end{equation}
To leading order in gravitational couplings, $a_A+a_B$ is of order $1/M_P$, while our results indicate that $a_A-a_B$ is of order $(m_A^2-m_B^2)/(M M_P^2)$. Hence, generically we expect the E\"otv\"os parameter to be of order 
\begin{equation}
	\eta_{AB}\sim \frac{m_A^2-m_B^2}{M M_P},
\end{equation}
which is negligible for practical purposes for elementary particle masses. But this does not necessarily rule out the phenomenological relevance of these corrections. If instead of using an on-shell renormalization scheme we had worked for instance with  minimal subtraction (MS), we would have found an E\"otv\"os parameter of order
\begin{equation}
	\eta\sim \frac{\mu^2}{M_\mu M_P^\mu}
	\left[\frac{(m_\mu^A)^4}{(m_I^A)^2}-\frac{(m_\mu^B)^4}{(m_I^B)^2}\right],
\end{equation}
where $m_\mu$ is the mass parameter in the MS scheme, and $m_I$ is the inertial mass. The key is that for light scalars (in the presence of fine tuning) the inertial mass $m_I$  may differ  from  the renormalized parameter $m_\mu \mu$ at a high scale $\mu\sim M_P$ by several orders of magnitude. In that case, the E\"otv\"os parameter may be of order one, and thus these quantum violations are phenomenologically relevant.  In any case, tests of the weak equivalence principle are not performed with elementary particles, but with macroscopic bodies instead.  In order to predict the  corresponding violations of the equivalence principle, we would have to proceed as in \cite{Damour:2010rm}.

\subsubsection{Ward-Takahashi Identity  for Broken Symmetry}

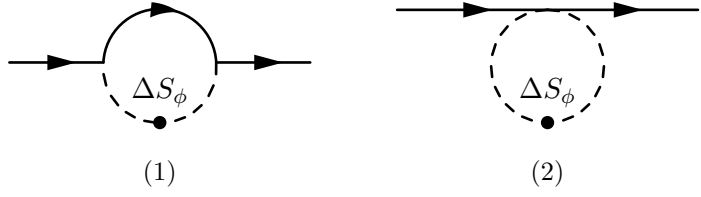
\begin{figure}
\subfigure[]
{
\begin{fmfgraph*}(40,16) 
\fmfleft{i1} 
\fmfright{o1}
\fmfbottom{vs} 
\fmf{fermion,tension=1.2}{v2,o1}
\fmf{fermion,tension=1.2}{i1,v1}
\fmf{fermion,left=.95}{v1,v2}
\fmffreeze
\fmf{dashes,left=0.45}{v2,vs,v1}
\fmfv{decor.shape=circle,decor.filled=full,decor.size=4,label=$\Delta S_\phi$,l.a=90}{vs}
\end{fmfgraph*}
}
\qquad 
\subfigure[]
{
\begin{fmfgraph*}(40,30) 
\fmfleft{i1} \fmfright{o1} \fmfbottom{vs} 
\fmf{fermion}{i1,vm,o1}
\fmffreeze
\fmf{dashes,left}{vm,vs}
\fmf{dashes,right}{vm,vs}
\fmfv{decor.shape=circle,decor.filled=full,decor.size=4,label=$\Delta S_\phi$,l.a=90}{vs}
\end{fmfgraph*}
}
\caption{Diagrams with two external lines and the insertion of the two vertices in equation (\ref{eq:delta S tot}).}
\label{fig:corrections}
\end{figure}

Our explicit calculation of the one-loop correction for scalar emission mediated by the scalar itself  also allows us to check the Ward-identity (\ref{eq:exact}) and   illustrate its meaning.  For that purpose let us rewrite equation (\ref{eq:exact}) in the form
\begin{equation}\label{eq:exact 2}
	\Gamma_\phi-\frac{1}{M}\Gamma_\Delta=\frac{2M_P}{M}(\Gamma_h)^\mu{}_\mu.
\end{equation}
On the left hand side of (\ref{eq:exact 2}), the corrections to $\Gamma_\phi$ to order $1/M^3$ are determined by the four diagrams in figure \ref{fig:scsv},  and are given by equations (\ref{eq:scalar delta vertex}). As we mention in Appendix \ref{sec:Ward identities for broken symmetries}, $\Gamma_\Delta$ is given by all 1PI diagrams with two external matter lines, and an insertion of the vertex $\Delta$, the change in the Lagrangian density under the transformation (\ref{eq:redef linear}). To calculate the sum of these diagrams to order  $1/M^3$ we just need to expand the change in the total action under the transformation (\ref{eq:redef linear}) to quadratic order in $\phi$. Since we are considering an exponential, equation (\ref{eq:f exp}), only $S_\phi$ changes under the transformation,
\begin{equation}
\label{eq:delta S tot}
	\Delta S_\mathrm{tot}=\frac{1}{2}\int d^d x\, 
	\left[(d-2)\partial_\mu \phi 
	\partial^\mu \phi+d\,  m_\phi^2 \phi^2	\right]\equiv \int d^d x \, \Delta.
\end{equation}
To leading order, insertion of this vertex in a diagram with two external lines then leads to  the two diagrams in figure \ref{fig:corrections}, which, respectively, contribute 
\begin{subequations}
\begin{align}
	\Gamma_{\Delta}^1&=-\frac{2i}{M^2} \int d^d k \, dx \, (1-x)
	\frac{[(d-2)(k+px)^2+d\, m_\phi^2][(d-2)p\cdot (p(1-x)-k)+d\, m^2]^2}
	{[k^2+p^2x(1-x)+m^2 x+m_\phi^2(1-x)]^3},\\
	\Gamma_{\Delta }^2&=\frac{i}{2M^2}\int d^d k\,
	\frac{[(d-2)k^2+d\, m_\phi^2][(d-2)^2 p^2+d^2\, m^2]}{[k^2+m_\phi^2]^2}.
\end{align}
\end{subequations}

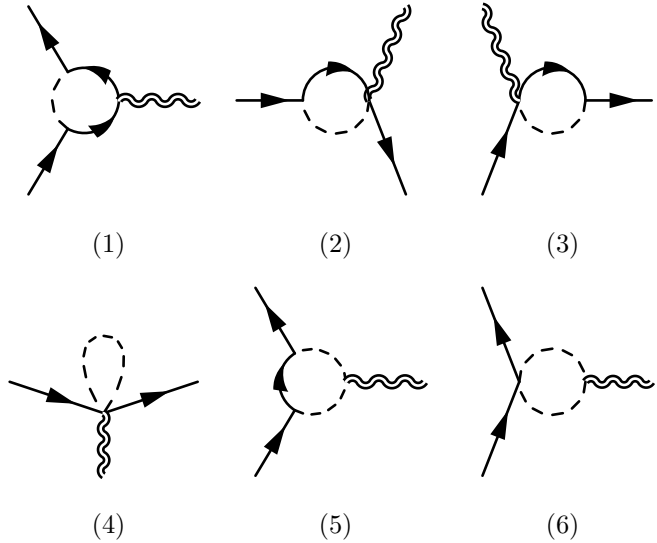
\begin{figure}
\renewcommand{\thesubfigure}{(\arabic{subfigure})}
\subfigure[]
{
\begin{fmfgraph}(25,25) 
\fmfleft{i1,i2} 
\fmfright{o1}
\fmf{fermion,tension=1.3}{i1,v1}
\fmf{fermion,tension=1.3}{v2,i2}
\fmf{dashes,left=.55}{v1,v2}
\fmf{fermion,right=.55}{v1,v3,v2}
\fmf{dbl_wiggly,tension=1.3}{v3,o1}
\end{fmfgraph}
}
\subfigure[]
{
\begin{fmfgraph}(25,25) 
\fmfleft{i1} 
\fmfright{o1,o2}
\fmf{fermion,tension=1.8}{v2,o1}
\fmf{fermion,tension=2}{i1,v1}
\fmf{fermion,left}{v1,v2}
\fmf{dashes,left}{v2,v1}
\fmf{dbl_wiggly,tension=1.8}{v2,o2}
\end{fmfgraph}
}
\subfigure[]
{
\begin{fmfgraph}(25,25) 
\fmfleft{i1,i2} 
\fmfright{o1}
\fmf{fermion,tension=1.8}{i1,v1}
\fmf{fermion,tension=2}{v2,o1}
\fmf{fermion,left}{v1,v2}
\fmf{dashes,left}{v2,v1}
\fmf{dbl_wiggly,tension=1.8}{v1,i2}
\end{fmfgraph}
}
\\ 
\subfigure[]
{
\begin{fmfgraph}(25,25) 
\fmfleft{i1} \fmfright{o1} \fmfbottom{s} 
\fmf{fermion}{i1,vs,o1}
\fmf{dashes,tension=.75}{vs,vs}
\fmf{dbl_wiggly}{vs,s}
\end{fmfgraph}
}
\subfigure[]
{
\begin{fmfgraph}(25,25) 
\fmfleft{i1,i2} 
\fmfright{o1}
\fmf{fermion,tension=1.3}{i1,v1}
\fmf{fermion,tension=1.3}{v2,i2}
\fmf{fermion,left=.55}{v1,v2}
\fmf{dashes,right=.55}{v1,v3,v2}
\fmf{dbl_wiggly,tension=1.3}{v3,o1}
\end{fmfgraph}
}
\subfigure[]
{
\begin{fmfgraph}(25,25) 
\fmfleft{i1,i2} 
\fmfright{o1}
\fmf{fermion,tension=1.8}{i1,v1,i2}
\fmf{dbl_wiggly,tension=2}{v2,o1}
\fmf{dashes,left}{v1,v2}
\fmf{dashes,left}{v2,v1}
\end{fmfgraph}
}
\caption{One-loop corrections to the vertex for graviton emission to order $1/M^2$.}
\label{fig:scgv}
\end{figure}

To calculate the right hand side of equation (\ref{eq:exact 2}) we need to expand the total action to first order in the graviton, and  second order in $(\phi/M)$,
\begin{multline}\label{eq:linear h}
	\mathcal{L}_\phi^E+\mathcal{L}^E_M\supset 
	-\frac{h_{\mu\nu}}{2M_P}
	\left[\eta^{\mu\nu}
	\left(\frac{1}{2}\partial_\rho \chi \partial^\rho \chi
	+\frac{1}{2}m^2\chi^2+
	\frac{1}{2}\partial_\rho\phi\partial^\rho\phi
	+\frac{1}{2}m^2\phi^2\right)-\partial^\mu \chi \partial^\nu \chi 
	-\partial^\mu \phi \partial^\nu \phi\right]-
	\\
	-\frac{h_{\mu\nu}}{2M_P}\frac{\phi}{M}
	\left[\eta^{\mu\nu}\left(\frac{d-2}{2}\partial_\rho\chi\partial^\rho\chi
	+\frac{d}{2}m^2\chi^2\right)-(d-2)\partial^\mu \chi \partial^\nu \chi\right]-
	\\
	-\frac{h_{\mu\nu}}{2M_P}\frac{\phi^2}{2M^2}
	\Bigg[\eta^{\mu\nu}\left\{\frac{(d-2)^2}{2}\partial_\rho\chi\partial^\rho\chi
	+\frac{d^2}{2} m^2\chi^2\right\}-(d-2)^2\partial^\mu \chi \partial^\nu \chi\Bigg].
\end{multline}
Then, to order $1/M^2$, $\Gamma_h$ on the right hand side of equation (\ref{eq:exact 2}) is given by the six diagrams in figure \ref{fig:scgv}, with vertices determined by the action (\ref{eq:linear h}). Let us label the contribution of the $i$-th diagram $(\Delta \gamma_i)^{\mu\nu}$. Then, we can write 
\begin{equation}
	(\Gamma_h)^{\mu\nu}=(\gamma_h)^{\mu\nu}
	+\sum_{i=1}^6 (\Delta\gamma_i)^{\mu\nu},
\end{equation}
where $(\gamma_h)^{\mu\nu}$ is the tree-level contribution.

Comparing the action (\ref{eq:scalar matter EF}) with (\ref{eq:linear h}) immediately reveals that the trace of the tree-level vertex for scalar emission by matter equals the trace of the tree-level vertex for graviton emission,
\begin{equation}\label{eq:check 0}
	\gamma_\phi=\frac{2M_P}{M}(\gamma_h)^\mu{}_\mu.
\end{equation}
This is just a reflection of the invariance of the matter action under (\ref{eq:redef linear}), as we discussed earlier. Therefore, it follows in addition that the contributions of diagram \ref{fig:scsv}.1, equation (\ref{eq:scalar delta vertex 1}), and the trace of that of \ref{fig:scgv}.1  are proportional to each other,
\begin{equation}\label{eq:check 1}
	\Delta \gamma_1=\frac{2M_P}{M}(\Delta\gamma_1)^\mu{}_\mu.
\end{equation}
Diagrams \ref{fig:scsv}.2 and \ref{fig:scsv}.3 are the same as those in \ref{fig:scgv}.2 and \ref{fig:scgv}.3.  In fact, since the quartic vertex in (\ref{eq:scalar matter EF}) is proportional to  the trace of the quartic vertex in equation (\ref{eq:linear h}), both pairs of diagrams basically yield identical contributions
\begin{equation}\label{eq:check 2}
	\Delta\gamma_2+\Delta\gamma_3=\frac{2M_P}{M}\left[(\Delta\gamma_2)^\mu{}_\mu
	+(\Delta\gamma_3)^\mu{}_\mu\right].
\end{equation}
Similarly, diagrams \ref{fig:scsv}.4 and \ref{fig:scgv}.4 are also identical, and because the quintic vertex in (\ref{eq:scalar matter EF}) is proportional to the trace of the quintic vertex in (\ref{eq:linear h}), both diagrams are again proportional to each other,
\begin{equation}\label{eq:check 3}
	\Delta\gamma_4=\frac{2M_P}{M}(\Delta\gamma_4)^\mu{}_\mu.
\end{equation} 
Furthermore, it is clear from the structure of the couplings  in (\ref{eq:scalar matter EF}) and (\ref{eq:linear h}) that these relations only apply for an exponential $f$. 

On the other hand, comparison of figures \ref{fig:scsv} and \ref{fig:scgv}   reveals that diagrams \ref{fig:scgv}.5 and \ref{fig:scgv}.6 do not have a scalar emission counterpart, simply because there is no analogous cubic vertex for   $\phi$ in the action. This is corrected for by  the two diagrams  with an insertion of $\Delta$ in figure \ref{fig:corrections}, whose contribution equals the trace of their graviton counterpart. To order $1/M^2$ this implies
\begin{equation}\label{eq:check 4}
	-\Gamma_{\Delta}=\frac{2M_P}{M}\left[(\Delta\gamma_5)^\mu{}_\mu+(\Delta\gamma_6)^\mu{}_\mu\right].
\end{equation}
Together,  equations (\ref{eq:check 0}), (\ref{eq:check 1}), (\ref{eq:check 2}), (\ref{eq:check 3}) and (\ref{eq:check 4}) immediately provide  an explicit verification of equation (\ref{eq:exact 2}).

We can further  test the validity of equation (\ref{eq:exact 2})  by noting that, because of equations (\ref{eq:h no corrections}) and (\ref{eq:check 0}), for fields renormalized on shell we should have
\begin{equation}
	(\Delta\gamma_\phi)_{OS}=\frac{\Gamma_{\Delta}}{M}.
\end{equation} 
Indeed, we have explicitly checked that in the limit $d\to 4$ both the pole and the finite parts on both sides of the last equation agree.   

\subsection{Fermion matter}

We turn our attention now to the vertex for scalar emission by fermionic matter. As we mentioned above, fermions are different from bosons because  coupling them to gravity necessarily requires the introduction of the vierbein. This section illustrates that, as far as the equivalence principle is concerned, this property does not introduce any new ingredients, and that the properties of the vertex with fermion matter closely resemble those of the vertex with scalar matter.

Consider the Jordan-frame matter action
\begin{equation}\label{eq:fermion EF L}
	S_M^J=\int d^d x \det e  \left[-\bar{\psi}e^\mu{}_a \gamma^a D_\mu \psi
	-m\bar{\psi}\psi-\frac{1}{2}\partial_\mu \chi \partial^\mu \chi
	-\frac{1}{2}m^2_\chi \chi^2-\lambda \chi \bar{\psi}\psi\right],
\end{equation}
which simply describes the Yukawa interactions of a spin 1/2 fermion $\psi$ with a massive Higgs-like scalar $\chi$. Here,  $\gamma^a$ are the conventional Dirac matrices and $D_\mu$ is the covariant derivative of the spinor, which depends on the vierbein through the spin connection. To obtain the Einstein frame action, we replace $e_\mu{}^a$ by $f(\phi/M) e_\mu{}^a$, and expand the resulting expression to the desired order in $\phi$ around flat space. But in order to calculate $S$-matrix elements, it is simpler to work with a Lagrangian in which some of the interactions have been removed by a field redefinition. It is well-known \cite{Birrell:1982ix} that the action of a massless spinor is invariant under the Weyl transformation
\begin{subequations}\label{eq:spinor conformal}
\begin{eqnarray}
	e_\mu{}^a&\to& f\, e_\mu{}^a, \\
	\psi&\to& f^{(1-d)/2}\,  \psi.
\end{eqnarray} 
\end{subequations}
Thus, making these substitutions in the Jordan-Frame action (\ref{eq:fermion EF L}) and expanding again to the required order we get in flat space
\begin{multline}\label{eq:fermion Lagrangian}
	\mathcal{L}_M^E=-\bar{\psi} \gamma^\mu \partial_\mu \psi
	-m\left(1+\frac{\phi}{M}
	+\frac{\phi^2}{2M^2}+\frac{\phi^3}{6M^3}\right)\bar{\psi}\psi
	-\lambda \chi \left(1+\frac{\phi}{M}\right)\bar{\psi}\psi-\\
	-\frac{1}{2}\left(1+(d-2)\frac{\phi}{M}\right)
	\partial_\mu \chi\partial^\mu \chi
	-\left(1+d\frac{\phi}{M}\right)\frac{1}{2}m_\chi^2 \chi^2
	+ \cdots,
\end{multline}
where we have assumed again that $f$ is an exponential, equation (\ref{eq:f exp}). Because  the couplings in this Lagrangian are not of the form (\ref{eq:S EF M}), the vertex amplitudes do not obey the Ward identity (\ref{eq:scalar Ward}), as can be easily verified at tree level. Instead,  because the field redefinition (\ref{eq:spinor conformal}) is of the form (\ref{eq:matter field redef}), the vertex obeys the dilatation Ward identity (\ref{eq:dilatation WI}), as can be also easily verified at tree level.   Note that in order to appropriately take into account the spinor field redefinition, we have to multiply the path integral measure by an appropriate Jacobian \cite{Brax:2010uq}. For an electrically neutral spinor, this has no effects to linear order in $\phi$.

\subsubsection{Matter Loops}

Our first goal is to calculate the order $\lambda^2/M$ corrections to the scalar-matter vertex induced by one-loop diagrams in which matter fields run inside the loop. The corresponding diagrams, figure \ref{fig:mcsv}, are the same as for scalar matter. We do not include external line corrections because we work in the OS scheme. In order to do so however, we need to introduce the appropriate counterterms  to enforce our renormalization conditions (\ref{eq:OS}). Introducing renormalized fields and mass parameters, $\psi\to Z^{1/2}_2\psi$ and $m\to m-\delta m$, we thus arrive at the counterterms
\begin{equation}
\mathcal{L}_C^E=-(Z_2-1)\left[\bar{\psi} \gamma^\mu \partial_\mu \psi+m\bar{\psi}\psi\right]+Z_2\delta m\bar{\psi}\psi
-\left[(Z_2-1)m-Z_2\delta m\right]\frac{\phi}{M}\bar{\psi}\psi+\cdots,
\end{equation}
where we have kept only those terms that are relevant for our calculation. 

The determination of the amplitudes associated with the diagrams in figure  \ref{fig:mcsv} is straight-forward. To simplify the analysis, we concentrate in the limit of zero momentum transfer and on-shell momenta, which is the appropriate limit for our considerations. Following the standard Feynman rules  (see e.g. \cite{Weinberg:1995mt})  we find that the contribution of the diagrams in figure \ref{fig:mcsv} is
\begin{subequations}\label{eq:fermion vc}
\begin{align}
	i\Delta\gamma_1&=-\frac{2\lambda^2m}{M}\int d^d k\int_0^1 dx \,
	\frac{x\left[m^2(2-x)^2-k^2\right]}{[k^2+m^2 x^2+ m_\chi^2(1-x)]^3},
	\label{eq:fermion vertex 1}\\
	i\Delta\gamma_2&=-\frac{2\lambda^2m}{M}\int d^d k\int_0^1 dx \,
	(1-x)\frac{(2-x)\left[(d-2)(k^2-m^2 x^2)+d m_\chi^2\right]-2(1-2/d)k^2 x}{[k^2+m^2 x^2+ m_\chi^2(1-x)]^3},
	\label{eq:fermion vertex 2}\\
	i\Delta\gamma_3&=\frac{\lambda^2 m}{M}\int d^d k\int_0^1 dx \,
	\frac{2-x}{[k^2+m^2 x^2+ m_\chi^2(1-x)]^2},
	\label{eq:fermion vertex 3} \\
	i\Delta\gamma_4&=i\Delta\gamma_3,
	\label{eq:fermion vertex 4}
\end{align}
\end{subequations}
where we have used that on shell we may substitute $\slashed{p}$ by $im$.

In addition, we need to consider the contributions of the counterterms, which in this case reduce to
\begin{equation}\label{eq:vertex correction c}
	i\Delta\gamma_5= -i (2\pi)^d (Z_2-1)\frac{m}{M}+i (2\pi)^d Z_2 \frac{\delta m}{M}.\\
\end{equation} 
We choose these counterterms to enforce the on-shell renormalization conditions (\ref{eq:OS}), which requires
\begin{equation}\label{eq:fermion OS}
	Z_2-1=-\frac{i}{(2\pi)^d}\frac{\partial \Delta\pi}{\partial\slashed{p}}\Big|_{\slashed{p}=im}, \,  \quad
	Z_2\delta m=-\frac{\Delta\pi(im)}{(2\pi)^d}.
\end{equation}
In order to calculate the values of the counterterms, we thus need to evaluate the self-energy correction. This is given by the diagram in figure \ref{fig:mse}, which finally leads to
\begin{subequations}\label{eq:fermion se}
\begin{align}
	i\Delta \pi &=\lambda^2 m \int d^d k\int_0^1 dx \,
	\frac{2-x}{[k^2+m^2 x^2+ m_\chi^2(1-x)]^2}, \\
	\frac{\partial \Delta \pi}{\partial \slashed{p}}&=
	-\lambda^2 \int d^d k\, dx \,
	\left[\frac{1-x}{[k^2+m^2 x^2+ m_\chi^2(1-x)]^2}
	+\frac{4m^2(2-x)(1-x)x}{[k^2+m^2 x^2+ m_\chi^2(1-x)]^3}\right].
\end{align} 
\end{subequations}
Then, the  total loop correction to the vertex for scalar emission by matter is
\begin{equation}\label{eq:fermion dgtot}
(\Delta \gamma_\phi)_{OS}=\sum_{i=1}^5 \Delta \gamma_i.
\end{equation}

According to the Ward identity (\ref{eq:scalar Ward}), the right hand side of equation (\ref{eq:fermion dgtot}) has to vanish, as the vertex correction only involves matter couplings in the loop. But as opposed to what happens in the scalar case, in which the  Ward identity at one-loop can be readily verified, one has to complete a surprising amount of work here  to show that $(\Delta\gamma_\phi)_{OS}$ equals zero. We leave this task for Appendix \ref{sec:Scalar Ward Identity for Fermions}, in which we explicitly prove that, indeed, 
\begin{equation}\label{eq:fermion cancellation}
	(\Delta \gamma_\phi)_{OS}=0,
\end{equation}
in agreement with our general result (\ref{eq:violation}). As before, the corresponding cancellation among the five different diagrams is an expression of diffeomorphism and Weyl invariance. 

If we regularize the theory by introducing a momentum cut-off $\Lambda$, diffeomorphism invariance is broken again, and the cancellation (\ref{eq:fermion cancellation}) does not hold. Instead, say, in the limit $m_\phi\to 0$ we find that $(\Delta\gamma_\phi)_{OS}$ is logarithmically divergent,
\begin{equation}\label{eq:fermion vertex OS}
	(\Delta\gamma_\phi)_{OS}\to \frac{7\lambda^2 \pi^2}{6}\frac{m}{M}-\lambda^2 \pi^2 \frac{m}{M} 
	\int_0^1 dx \, (5-14x+6x^2)\log \frac{\Lambda^2}{m^2 x^2}.
\end{equation}
As in the scalar case, in order to renormalize this divergence we would have to introduce a coupling constant counterterm $\delta M^{-1}$ to the Lagrangian, which would contribute
\begin{equation}
	i\Delta\gamma_6=-i (2\pi)^4 \delta M^{-1} m
\end{equation}
to the vertex amplitude. In that case, we could \emph{impose} the condition $(\Delta\gamma_\phi)_{OS}+\Delta\gamma_6=0$, which would guarantee the preservation of the weak equivalence principle at one loop. But of course, since neither the Yukawa coupling $\lambda$ nor the mass $m$  are universal, this would lead to a collection of widely different set of bare coupling constants $M_i$, one for each fermion species, and it would remain a mystery why the renormalized vertex correction for all of them vanishes \ at zero momentum transfer. Otherwise, equation (\ref{eq:fermion vertex OS}) implies generic values of the E\"otv\"os parameter $\eta_{AB}$ of order $\lambda^2$.

\subsubsection{Scalar Loops}
Having seen how matter loop corrections do respect the equivalence principle (in the dimensionally regularized theory), let us turn our attention to those  corrections that do lead to violations. This time, instead of looking at diagrams with matter loops, we shall calculate the corrections caused by a scalar field loop, at order $1/M^3$. 

The one-loop scalar field corrections to the scalar vertex are the same as for scalar matter; they are shown in figure \ref{fig:scsv}. The self-energy corrections are also given by the  diagrams in figure \ref{fig:sse}. Comparison of the corrections to the vertex caused by a fermion loop to those caused by the scalar shows that vertices and most of the diagrams basically agree if one replaces fermion lines with scalar lines. Therefore, we can borrow the results of the previous subsection, now keeping the momenta off-shell, simply by replacing $\lambda$ by $m/M$, and $m_\chi^2$ by $m_\phi^2$. We do not need to consider the contribution of equation  (\ref{eq:fermion vertex 2}),   which does not have a counterpart in the scalar loop diagrams at order $1/M^3$.  Therefore, the  vertex  loop correction is the sum of the four terms
\begin{subequations}
\begin{align}
	&i\Delta\gamma_1=\frac{2m^3}{M^3}\int d^d k\int_0^1 dx \,x\,
	\frac{p^2(1-x)^2+2i\slashed{p}\,m(1-x)+k^2-m^2}
	{[k^2+p^2x(1-x)+m^2 x+ m_\phi^2(1-x)]^3},
	\\
	&i\Delta\gamma_2=\frac{m^2}{M^3}\int d^d k\int_0^1 dx \,
	\frac{-i\slashed{p}(1-x)+m}{[k^2+p^2x(1-x)+m^2 x+ m_\phi^2(1-x)]^2},
	\\
	&i\Delta\gamma_3=i\Delta\gamma_2,
	\\
	&i\Delta\gamma_4=-\frac{m}{2M^3}\int d^d k \, \frac{1}{k^2+m_\phi^2}.
\end{align}
\end{subequations}

The contribution from the counterterms is still given by (\ref{eq:vertex correction c}), with the latter determined by equations (\ref{eq:fermion OS}). But  this time, there is a new contribution to the self-energy, captured by the second diagram in figure \ref{fig:sse},
\begin{subequations}
\begin{align}
	&i\Delta \pi_1=\frac{m^2}{M^2} \int d^d k\int_0^1 dx \,
	\frac{-i\slashed{p}(1-x)+m}{[k^2+p^2x(1-x)+m^2 x+ m_\phi^2(1-x)]^2}, 
	\\
	&i\Delta \pi_2=-\frac{m}{2M^2}\int d^d k \, \frac{1}{k^2+m_\phi^2}.
\end{align}
\end{subequations}
In this case, when we add the contributions from of order $(m/M)^3$, we find that the cancellations that occurred at order $\lambda^2/M$ before do not operate.  To actually see that the overall vertex correction $(\Delta\gamma_\phi)_{OS}$ indeed is  different from zero, let us consider again the limit $d\to 4$. In this limit, the correction approaches
\begin{equation}\label{eq:fermion pole}
	(\Delta\gamma_\phi)_{OS}=\frac{2\pi^2}{M^3}\frac{i m^2\slashed{p}-2m^3}{d-4}+\mathcal{O}[(d-4)^0],
\end{equation}
which again  shows that the theory defined by (\ref{eq:S JF}) is not renormalizable, in the  sense that we cannot absorb this pole by 
renormalization of the coupling constant $M^{-1}$  in the Lagrangian (\ref{eq:fermion Lagrangian}). As in the scalar case, once this pole is removed by including  the appropriate missing counterterms in the action, we expect then finite vertex corrections of order $m^3/M^3$ in the limit $d\to 4$, which lead to relative violations of the weak equivalence principle of order $m^2/M^2$.  

\subsubsection{Ward-Takahashi Identity  for Broken Symmetry}

We mentioned in Section \ref{sec:Extension of the Conformal Symmetry to the Full Action} that in flat spacetime, scalar-tensor theories posses a broken dilatation symmetry (\ref{eq:dilatation}), and  a corresponding Ward identity for this broken symmetry, equation (\ref{eq:dilatation WI}). Again, we can use the results of our explicit calculation of the vertex correction in the previous section to check the validity of the Ward identity (\ref{eq:dilatation WI}), and vice-versa.

Since a fermion has scaling dimension $(d-1)/2$, the vertex for scalar emission $\Gamma_\phi$ by fermion matter obeys the identity (\ref{eq:dilatation WI}) with $n=1$. The Lagrangian (\ref{eq:fermion Lagrangian}) is not invariant under the dilatation (\ref{eq:dilatation}), but instead changes  by equation (\ref{eq:delta S tot}).  Therefore,  we should have
\begin{equation}\label{eq:fermion conformal WI}
	M \, \Gamma_\phi
	+ \left(p^\mu \frac{\partial}{\partial p^\mu}-1\right)
	\Pi=\Gamma_\Delta,
\end{equation}
where $\Gamma_\Delta$ is  the sum of all 1PI diagrams with two external lines and an insertion of the local operator $\Delta$ defined in equation (\ref{eq:delta S tot}). Recall that the Ward identity (\ref{eq:exact}) does not hold in this case, as can be readily verified at tree level, because the field redefinition (\ref{eq:spinor conformal}) has led to a matter action that is not of the form (\ref{eq:S EF M}).

At tree level, it is easy to check the validity of (\ref{eq:fermion conformal WI}), since there is no tree-level diagram with an insertion of $\Delta$ and two fermion lines. At order $1/M^2$, the corresponding Feynman diagrams are those in figure \ref{fig:corrections}, which yield the two correction terms
\begin{subequations}
\begin{align}
	&\Gamma_\Delta^1=-\frac{2im^2}{M^2}
	\int d^d k \, dx \, (1-x)\,
	\frac{\left[-i\slashed{p}(1-x)+m+i\slashed{k}\right]
	\left[(d-2)(k+x\,p)^2+d\,m_\phi^2\right]}
	{[k^2+p^2x(1-x)+m^2x+m_\phi^2(1-x)]^3},\\
	& \Gamma_\Delta^2=\frac{im}{2M^2}
	\int d^dk\, \frac{(d-2)k^2+d\,m_\phi^2}{[k^2+m_\phi^2]^2}.
\end{align}
\end{subequations}
It is then easy to check for instance that the residue of the pole at $d=4$ in $\Gamma_\Delta\equiv\Gamma_\Delta^1+\Gamma_\Delta^2$ actually agrees with equation (\ref{eq:fermion pole}), thus confirming the validity of the Ward identity (\ref{eq:fermion conformal WI}).

\section{Summary and Conclusions}

We have studied the impact of quantum corrections on the  weak equivalence principle in scalar-tensor theories that admit a Jordan-frame formulation, equation (\ref{eq:S JF}). To do so, it is convenient to work in the Einstein frame, in which the scalar and the graviton are decoupled in the free action, equation (\ref{eq:S EF}). In this frame the amplitude for scalar emission is universally proportional to the inertial mass at tree level, and the same result holds when we include quantum corrections that only involve matter loops. Once we include a scalar $\phi$ or a graviton in these loop corrections however, the equivalence principle is violated.  

The origin of these results lies in the broken Weyl symmetry (\ref{eq:redef linear}). The corresponding Ward identity for the broken symmetry (\ref{eq:exact}) relates the 1PI vertex $\Gamma_\phi$ for scalar emission to that of the graviton $\Gamma_h$, and to the sum of all the diagrams with an insertion of a vertex proportional to the change of the Lagrangian density under the broken symmetry (\ref{eq:exact}), $\Gamma_\Delta$. Violations of the equivalence principle caused by the scalar interaction arise from those terms in the action that violate the shift symmetry (\ref{eq:redef linear}). For an exponential, $f=\exp(\phi/M)$, the matter action is exactly symmetric under (\ref{eq:redef linear}) and only $S_\phi$ and $S_{EH}$ violate the Weyl symmetry. For other choices of $f$, such as a linear coupling in $\phi$, even the matter Lagrangian is not exactly symmetric under this transformation. In both cases, because the only terms that violate the inhomogeneous Weyl symmetry involve terms quadratic in the scalar $\phi$ or the graviton, these violations of the equivalence principle are proportional to three powers of the gravitational couplings $M^{-1}$ and $M_P^{-1}$. If we regularize the theory with a momentum cut-off, diffeomorphism invariance is broken, and even matter loops lead to violations of the weak equivalence principle caused by the scalar interaction. Although diffeomorphism invariance is required to couple a massless graviton to matter, there is no analogous constraint to couple a massive or massless scalar to matter. In particular, a momentum cut-off does break the Weyl symmetry  (\ref{eq:redef linear}), but the latter is broken anyway in the action (\ref{eq:S EF}). 

The form of  the quantum corrections to the scalar vertex $\Gamma_\phi$ implies that scalar-tensor theories with an Einstein frame formulation of the form (\ref{eq:S EF M}) are not renormalizable: Any matter action  of the form (\ref{eq:S EF M}) does not contain enough counterterms to eliminate all the poles at $d=4$ in the dimensionally regularized theory. To do so one has to include all the terms compatible with the symmetries of the action, which only consist of diffeomorphism invariance. Therefore, the structure of (\ref{eq:S EF M}) is not preserved by quantum corrections. From that point of view, assuming that the coupling of the scalar is universally characterized by a single coupling constant $1/M$ appears artificial.

The actual magnitude of the equivalence principle violations depends on the way the theory is regularized, and on the renormalization prescription that eliminates the remaining non-renormalizable divergences in the amplitudes.  Generically, in the presence of a high momentum cut-off, we expect the E\"otv\"os parameter of these theories to be of order one, which is strongly ruled out by experiment \cite{Schlamminger:2007ht}. In the dimensionally regularized theory,  we expect the E\"otv\"os parameter to be of order $\Delta m^2/M_P^2$; this ratio is extremely small for typically inertial masses of elementary particles, but could be large if one of the mass parameters is defined away from the mass shell. In any case, we have not worked out the magnitude of the equivalence principle violations for macroscopic bodies, as appropriate  for phenomenological considerations. 

Finally, our results can be easily extended to similar classes of theories in which the matter action can be cast as in equation (\ref{eq:S EF M}), such as $f(R)$ gravity \cite{Carroll:2003wy} or the Galileon \cite{Nicolis:2008in}. Because both of them violate the Weyl symmetry (\ref{eq:redef linear}), we expect them  to behave like the scalar-tensor theories we have considered here.

\acknowledgments
We thank Alberto Nicolis and Eanna Flanagan for useful conversations and feedback. This work is supported in part by the NSF Grant PHY-0855523. 

\appendix
\section{Ward-Takahashi Identities for Broken Symmetries}
\label{sec:Ward identities for broken symmetries}
It is well-known that linear symmetries of the action are also symmetries of the effective action.  In this appendix we are concerned with transformations that, though linear, do not preserve the form of the action. As we shall show, in this case, the quantum effective action satisfies a Ward-Takahashi identity that relates the change of the effective action under the  linear transformation to the change in  total action functional under the broken symmetry. This general identity has been widely discussed in the literature, see e.g. \cite{Coleman}, though its proof is difficult to find. Our derivation here closely follows the formalism of \cite{ZinnJustin:1993wc} (particularly its Section 12.6).

Consider  the generating functional of an arbitrary theory that contains a set of fields $\chi^n$ in the presence of a corresponding set of currents $J_n$,
\begin{equation}
	Z[J]=\int D\chi 
	\exp\left(i S_\mathrm{tot}[\chi]
	+i\int d^dx \, J_n(x) \chi^n(x) \right).
\end{equation} 
Suppose now that we change integration variables
\begin{equation}\label{eq:linear transformation}
\chi^n(x)\to \chi^n(x)+\epsilon \Delta\chi^n(x),
\end{equation}
where $\Delta\chi^m$ is \emph{linear} in the fields, and $\epsilon$ is an arbitrary infinitesimal constant that we use as an expansion parameter (the actual transformation (\ref{eq:linear transformation}) may be global or local). Then, invariance of the path integral under change of variables  gives, to first order in $\epsilon$,
\begin{equation}\label{eq:step 1}
	\int D\chi
	\left(\Delta S_\mathrm{tot}[\chi]+\int d^d x \, J_n(x)\Delta\chi^n(x)\right)
	\exp\left(i S_\mathrm{tot}[\chi]
	+i \int  d^dy \, J_n(y) \chi^n(y)\right)=0,
\end{equation}
where $\Delta S_\mathrm{tot}$ is the total change in the action under the transformation (\ref{eq:linear transformation}), and we have also absorbed an eventual change  of the functional measure into $\Delta S_\mathrm{tot}$. 

In order to take into account the change of the action under the transformation, it turns out to be convenient to introduce a new generating functional  $Z[J,B]$ with an additional (constant) source $B$ for $\Delta S_\mathrm{tot}$,
\begin{equation}
Z[J,B]\equiv\int D\chi 
	\exp\left(i S_\mathrm{tot}[\chi]
	+i\int d^dx \, J_n(x) \chi^n(x)+i \,B \, \Delta S_\mathrm{tot}[\chi] \right).
\end{equation}
Then, in terms of this new functional, equation (\ref{eq:step 1}) takes the form 
\begin{equation}\label{eq:step 2}
	\left(\frac{1}{i\, Z[J,B]} \frac{\partial Z[J,B]}{\partial B}
	+\int d^d x \, J_n(x)\, \langle \Delta\chi^n(x)\rangle_{J, B}\right)\Bigg|_{B=0}=0,
\end{equation}
where, for any functional $F[\chi]$ of the fields, we have defined
\begin{equation}
	\langle F[\chi]\rangle_{J, B}\equiv 
	Z^{-1}[J,B] \int D\chi\,  F[\chi]
	\exp\left(i S_\mathrm{tot}[\chi]+i \int  d^dx \, J_n(x) \chi^n(x)
	+i  \, B \, \Delta S_\mathrm{tot}[\chi] \right).
\end{equation}
 
We proceed now to turn Equation (\ref{eq:step 2}) into an equation for the effective action. We first define the generating function for connected diagrams in the presence of a source for $\Delta S_\mathrm{tot}$ in the standard way,
\begin{equation}
	i\,  W[J,B]\equiv \log Z[J,B],
\end{equation}
and then  introduce the effective action by a Legendre transformation that only involves the curents $J_n$,
\begin{equation}\label{eq:gamma J B}
	\Gamma[\bar{\chi},B]\equiv W[J[\bar{\chi},B],B]-\int d^dx \, J_n[\bar{\chi},B] \bar{\chi}^n.
\end{equation}
The currents $J[\bar{\chi},B]$ in the last equation are such that the fields $\chi^n$ have prescribed expectation values\footnote{If the generating functional depends on the fields $\chi$, and some background values $\bar{\chi}$ through  gauge-fixing and ghost terms, the effective action is a functional of both  the background fields $\bar{\chi}$ and the expectation values of the fields  in the presence of the current, $\Gamma=\Gamma[\bar{\chi},\langle\chi\rangle_J]$ (we set here $B=0$ for simplicity.) The effective action in the background field method is defined by setting $\langle\chi\rangle=\bar\chi$, so, strictly speaking, the proper vertices are given by   functional derivative of $\Gamma=\Gamma[\bar{\chi},\langle\chi\rangle]$ with respect to $\langle \chi^n\rangle$. As shown in \cite{Abbott:1983zw} however, the difference is irrelevant when computing $S$-matrix elements.}
 $\bar{\chi}^n(x)$,
\begin{equation}\label{eq:effective definition}
	\langle \chi^n (x)\rangle_{J, B}=\frac{\delta W[J,B]}{\delta J_n(x)}=\bar{\chi}^n(x).
\end{equation}
Therefore, differentiation of equation (\ref{eq:gamma J B}) with respect to $\bar{\chi}^n$ and $B$ respectively  leads to the identities
\begin{equation}\label{eq:diff identities}
	J_n[\bar{\chi},B]=-\frac{\delta \Gamma[\bar{\chi},B]}{\delta \bar{\chi}^n}, \quad\quad 
	\frac{\partial \Gamma[\bar\chi,B]}{\partial B}=\frac{\partial W[J[\bar{\chi},B],B]}{\partial B}\Bigg|_J.
\end{equation}

We are ready to put all these results together into equation (\ref{eq:step 2}). First, note that the first term on the left-hand side is simply the derivative of $W[J,B]$ with respect to $B$, which, because of (\ref{eq:diff identities}) equals the derivative of the effective action $\Gamma[J,B]$ with respect to the same variable. Moreover, because we assume that $\Delta\chi^n$ is linear in the fields, 
\begin{equation}\label{eq:delta linear}
	\langle \Delta\chi^n\rangle_{J, B}=\Delta\bar\chi^n,
\end{equation}
so  the second term on the left-hand side of equation (\ref{eq:step 2}) is the change in the effective action $\Delta\Gamma[\bar{\chi}]\equiv\Gamma[\bar{\chi},0]$ under the transformation (\ref{eq:linear transformation}).  Therefore, equation (\ref{eq:step 2}) reads
\begin{equation}\label{eq:pre WI}
	\Delta\Gamma[\bar{\chi}]
	=\frac{\partial \Gamma[\bar{\chi}, B]}{\partial B}\Bigg|_{B=0},
\end{equation}
which states that at $B=0$ the effective action $\Gamma[\bar{\chi},B]$ is invariant under the transformation (\ref{eq:linear transformation}), supplemented with the additional transformation $B\to B-\epsilon$. 

The right-hand side of equation (\ref{eq:pre WI}) has a simple interpretation. Typically, $\Delta S_\mathrm{tot}$ is the spacetime integral of a local operator,\footnote{Care should  be exercised here because we are discarding  a surface terms that may arise upon integration by parts when isolating the change in the action to first order in $\epsilon$.}
\begin{equation}\label{eq:delta def}
	\Delta S_\mathrm{tot}=\int d^d x\, \Delta.
\end{equation}
In that case, $\Gamma[\bar\chi,B]$ is  the generator of 1PI diagrams in a theory with an additional interaction $\int d^d x\, B\, \Delta$. Therefore, its derivative with respect to the ``coupling constant" $B$ at zero simply picks up those 1PI diagrams with a single insertion of the vertex $\Delta$.  Since the new interaction involves a spacetime integral, and $B$ is a constant, such an insertion  carries zero momentum into the diagram.
Thus, denoting by $\Gamma_\Delta[\bar\chi]\equiv (\partial \Gamma/\partial B)\big|_{B=0}$ the generator of all 1PI diagrams with a vertex insertion of $\Delta$ we arrive at the main result of the appendix, 
\begin{equation}\label{eq:WI}
	\Delta\Gamma[\bar{\chi}]
	=\Gamma_\Delta[\bar\chi],
\end{equation}
the  Ward identity for a broken symmetry expressed in terms of the effective action (variations of the same identity are also known as Slavnov-Taylor or Schwinger-Dyson equations.)  By taking functional derivatives of equation (\ref{eq:WI}) with respect to the matter fields  one can then derive relations between the 1PI vertices of the theory. For instance,
\begin{equation}
	\Gamma^{\beta\alpha}_\Delta\equiv \frac{\delta^2 \Gamma_\Delta}{\delta\bar \psi_\alpha(x) \delta \bar\psi^\dag_\beta(y)}\Bigg|_{\bar\psi=0}
\end{equation} 
is the sum of all 1PI diagrams with two external matter fields (with propagators stripped off) and an insertion of $\Delta$.  Note that if the theory is invariant under the transformation (\ref{eq:linear transformation}), $\Delta=0$, equation (\ref{eq:WI}) reduces to the well-known Slavnov-Taylor identity for a linear symmetry of the action.

\section{Scalar Ward Identity for Fermions}
\label{sec:Scalar Ward Identity for Fermions}
In this appendix we verify that matter loop corrections do not renormalize the vertex for scalar emission by a fermion, equation (\ref{eq:fermion cancellation}).

With a  scalar $\chi$ running inside the loop, the one-loop correction to the vertex for scalar emission by a fermion  is determined by equations (\ref{eq:fermion vc}) and (\ref{eq:vertex correction c}), whereas the one-loop correction to the self-energy of the fermion is given by equations (\ref{eq:fermion se}). To verify the relation (\ref{eq:fermion cancellation}) we need to explicitly carry out the integrals over momenta and $x$. 

The integrals over momenta can be easily performed using the identity (\ref{eq:momentum int}). On shell, the remaining integrals over $x$ turn out to be  a sum of expressions of the general form
\begin{equation}\label{eq:In}
	I_{n}=\int_0^1 dx \,  x^n \left[m^2 x^2+m_\chi^2(1-x)\right]^{d/2-3},
\end{equation}
with integer $n$. After completing a square inside the square bracket,  the integral can be re-expressed as
\begin{equation}
	I_n=\int_0^1  dx\,  x^n (m_\chi)^{d-6}
	\left(1-\frac{1}{4r}\right)^{d/2-3}
	\left[1+\frac{r}{1-\frac{1}{4r}}
	\left(x-\frac{1}{2r}\right)^2\right]^{d/2-3},
\end{equation}
where we have defined the dimensionless ratio
\begin{equation}
	r\equiv\frac{m^2}{m_\chi^2}.
\end{equation}
The scalar $\chi$ is stable upon decay onto two fermions  if $m_\chi<2 m$. In that case $(1-1/4r)>0$, and we can introduce the new (real) integration variable
\begin{equation}\label{eq:t}
	t=\sqrt{\frac{4r^2}{4r-1}}\left(x-\frac{1}{2r}\right).
\end{equation}
In terms of this variable, the integral (\ref{eq:In}) thus becomes
\begin{equation}\label{eq:t integral}
	I_{n}= m_\chi^{d-6} 
	\left(1-\frac{1}{4r}\right)^{d/2-3}\sqrt{\frac{4r-1}{4r^2}	}
	\int_{t_0}^{t_1} dt\, \left(\sqrt{\frac{4r-1}{4r^2}}\,t+\frac{1}{2r}\right)^n
	\left[1+t^2\right]^{d/2-3},
\end{equation}
where the lower and upper integration limits $t_0$ and $t_1$ are  determined by respectively setting $x=0$ and $x=1$ in equation (\ref{eq:t}). Expanding the $n$-th power in equation (\ref{eq:t integral}) we further obtain a linear combination of integrals of the general form
\begin{equation}
	J_{m}\equiv \int_{t_0}^{t_1} dt\, t^m \left[1+t^2\right]^{d/2-3},
\end{equation}
which can finally be expressed in terms of hypergeometric functions \cite{Abramowitz},
\begin{equation}
J_{m}=\frac{t_1^{1+m}}{1+m}\,
	{}_2F_1\left(3-\frac{d}{2},\frac{1+m}{2},\frac{3+m}{2};-t_1^2\right)
	-\frac{t_0^{1+m}}{1+m}\,
	{}_2F_1\left(3-\frac{d}{2},\frac{1+m}{2},\frac{3+m}{2};-t_0^2\right).
\end{equation}
In this way, after quite a bit of tedious but straight-forward algebra, collecting all the contributions from the integrals in equation (\ref{eq:fermion dgtot}) we find that they all add to zero, equation (\ref{eq:fermion cancellation}).

\end{fmffile}

\end{document}